\documentclass[sigconf]{acmart}
\usepackage{pgfplots}
\pgfplotsset{compat=1.18}
\pdfoutput=1
\usepackage[utf8]{inputenc}
\usepackage[english]{babel}
\usepackage[T1]{fontenc}
\usepackage{amsmath}
\usepackage{hyperref}

\usepackage{tikz}
\usepackage{lipsum}

\AtBeginDocument{%
  }

\usepackage[utf8]{inputenc}
\usepackage{pgfplots}
\pgfplotsset{compat=newest}
\usepgfplotslibrary{statistics}

\usepackage{color}
\definecolor{Diagramm1}{HTML}{636a6e}
\definecolor{Diagramm2}{HTML}{227d41}
\definecolor{Diagramm3}{HTML}{56be79}
\definecolor{Diagramm4}{HTML}{b3e3c3}
\definecolor{Diagramm5}{HTML}{969696}
\definecolor{Diagramm6}{HTML}{d9d9d9}
\definecolor{Diagramm7}{HTML}{3b4042}

\usepackage{multirow}%
\usepackage{amsmath,amsfonts}%
\usepackage{mathrsfs}%
\usepackage[title]{appendix}%
\usepackage{textcomp}%
\usepackage{manyfoot}%
\usepackage{booktabs}%
\usepackage{algorithm}%
\usepackage{algorithmicx}%
\usepackage{algpseudocode}%
\usepackage{listings}
\usepackage{comment}

\usepackage{cleveref}
\usepackage{tikz}
\usetikzlibrary{decorations.text,patterns,shapes.geometric,positioning,shapes.multipart,backgrounds,calc,fit}

\usepackage[utf8]{inputenc}
\usepackage{pgfplots,pgfplotstable}\pgfplotsset{compat=newest}
\usepgfplotslibrary{fillbetween}
\usepackage{caption}
\usepackage{subcaption}
\usepackage{makecell}
\usepackage{listings}
\usepackage{todonotes}

\theoremstyle{thmstyleone}%

\theoremstyle{thmstyletwo}%

\theoremstyle{thmstylethree}%

\usepackage{listings}

\lstdefinelanguage{SQL}{
    keywords={SELECT, FROM, WHERE, JOIN, ON, ORDER BY, GROUP BY, ASC, DESC, LIMIT, SUM, MIN},
    keywordstyle=\color{blue}\bfseries,
    comment=[l]{--},
    morecomment=[s]{/*}{*/},
    commentstyle=\color{gray}\ttfamily,
    stringstyle=\color{red}\ttfamily,
    identifierstyle=\color{black},
    basicstyle=\ttfamily\footnotesize
}

\begin{document}
\title{Improved Join Order Optimization for Database Queries using Hybrid Quantum-Classical Approaches for QUBO Problems}

\author{Nitin Nayak}
\thanks{Both authors contributed equally to this research.}
\email{nitin.nayak@uni-luebeck.de}
\orcid{0009-0007-4015-8602}
\affiliation{%
  \institution{Universität zu Lübeck}
  \city{Lübeck}
  \state{Schleswig-Holstein} 
  \country{Germany}
}

\author{Tobias Winker}
\affiliation{%
  \institution{Universität zu Lübeck}
  \city{Lübeck}
  \state{Schleswig-Holstein}
  \country{Germany}}
\email{t.winker@uni-luebeck.de}

\author{Umut Çalıkyılmaz}
\affiliation{%
  \institution{Universität zu Lübeck}
  \city{Lübeck}
  \state{Schleswig-Holstein}
  \country{Germany}}
\email{umut.calikyilmaz@uni-luebeck.de}

\author{Jinghua Groppe}
\affiliation{%
  \institution{Universität zu Lübeck}
  \city{Lübeck}
  \state{Schleswig-Holstein}
  \country{Germany}}
\email{jinghua.groppe@uni-luebeck.de}

\author{Sven Groppe}
\authornotemark[1]
\email{sven.groppe@informatik.tu-freiberg.de}
\affiliation{%
  \institution{TU Bergakademie}
  \city{Freiberg}
  \state{Saxony}
  \country{Germany}
}

\renewcommand{\shortauthors}{Nayak et al.}

\begin{abstract}
    Efficient query optimization is crucial for relational database systems, especially for optimizing join orders in complex queries. This work introduces a hybrid approach that integrates Eliminating Cartesian Products (ECP) with splitting the QUBO search space (SQSS) to reduce the size of the QUBO problem, minimizing binary variables and constraints. This improves the performance of the quantum algorithm while lowering hardware requirements.
    We evaluate our method using real-world SQL queries from the ErgastF1 dataset on quantum and classical algorithms, including Quantum Annealing (QA), Simulated Annealing (SA), QAOA, and VQE, implemented on D-Wave’s Quantum Annealer and universal gate-based simulators. 
    Additionally, we analyze the impact of selectivity and SQSS on QUBO weight distribution and algorithmic performance, highlighting optimization efficiency for QA and SA.
    Experimental results show consistent optimal join orders and enhanced query optimization for various selectivity conditions, and it also highlights the limitations of current quantum hardware for complex queries. This study further confirms the potential of hybrid quantum-classical methods for scalable quantum-enhanced database optimization.
\end{abstract}



\maketitle

\section{Introduction}
\label{sec:introduction}

Join order plays a key role in database query performance and join order optimization (JOO), which aims at determining the optimal sequence of joins, and has therefore been an active research topic. The JOO problem is known to be an NP-hard combinatorial optimization problem~\cite{scheufele1996constructing}, and as the number of relations increases, it becomes a challenging task to develop efficient algorithms to find good enough solutions in a reasonable amount of time. Dynamic programming algorithms search all possible combinations of join sequences and so can deliver optimal solutions but they suffer from an exponential computing complexity to the number of relations \cite{warnke2024rejoosp,astrahan1976system,graefe1995cascades,graefe1993volcano,lamb2012vertica,soliman2014orca}. Instead of exploring all possible permutations, heuristic approaches use pre-defined rules and strategies that prioritize promising join sequences to compute a reasonably good solution in an acceptable time \cite{warnke2024rejoosp, winslett2020goetz, moerkotte2006analysis, waas2000join}. However, such approaches do not guarantee the optimal join orders, potentially leading to suboptimal performance. Furthermore, the quality of the optimization heavily depends on the heuristics used, and some heuristics may work well for certain types of queries or data distributions but fail in others. Instead of manually defining rules and strategies, machine learning approaches \cite{warnke2024rejoosp, marcus2019neo, zhang2020alphajoin, park2020quicksel, marcus2018deep, krishnan2018learning, ji2023query} try to learn them from experiences. Once trained, machine learning models can find a solution very quickly. However, the optimization quality of ML models heavily relies on the training datasets, and they also adapt poorly to changes in data distribution or query patterns.

Quantum computing provides a new and promising solution to the JOO. Due to the unique properties of quantum mechanics, quantum computing devices have the ability to process a large number of potential solutions at the same time, rather than exploring each possibility in turn like traditional computers do, and this ability is especially beneficial for solving combinatorial optimization problems. One approach to solving a combinatorial optimization with a quantum computer is the formulation as a quadratic unconstrained binary optimization (QUBO) problem. As quantum computing devices are becoming increasingly accessible, the database community is trying to apply the ability of quantum computing to the JOO problem. The work in \cite{schonberger2023ready} suggested a QUBO solution for the JOO, but it restricts its solution spaces to left-deep trees and can not process general bushy joins. \cite{schonberger2023ready} also observed that current-stage quantum processing units (QPU) can only optimize small-scale queries and cannot yield meaningful results when optimizing more complex join queries. The work \cite{schonberger2022quantum} proposes a quantum QUBO-encoding schema with indirect join cost for the join ordering problem to process general bushy join trees, but it has not evaluated the technique. Furthermore, the adoption of indirect join costs could lead to suboptimal solutions. 

To address the challenges mentioned above, the work in \cite{nayak2023constructing} proposes a QUBO formulation with direct join cost for JOO to process general bushy join trees, and the techniques are proven to have the best computational complexity that can be achieved theoretically. The authors in \cite{nayak2023constructing} also experimentally evaluate their techniques, and the evaluation results show that these techniques achieve better performance in finding valid and optimal shots for real-world queries than the work in \cite{schonberger2023ready}. To mitigate the hardware limitations of current QPUs as reported in \cite{schonberger2023ready}, the authors in \cite{nayak2024quantum} extend the work in the paper \cite{nayak2023constructing} with a splitting technique that partitions a big search space of QUBO problems into smaller subspaces and thus enables QPUs to process more complex join optimizations. 

Our paper extends the work in \cite{nayak2024quantum} with the following new contributions to further improve the quantum-based join order optimization: 
\begin{itemize}
\item a technique of eliminating the cartesian product to reduce the search space and intermediate join results. 
\item an extensive experimental evaluation, which evaluates the existing JOO algorithms (Dynamic Programming, Simulated Annealing, Quantum Annealing, Variational Quantum Eigensolver, and Quantum Approximation Optimization Algorithm) integrated with our techniques and compares them with the original JOO algorithms on different quantum computing devices
(gate-based quantum simulators, and D-Wave’s quantum annealer)
\item a first work that investigates how different selectivities in queries impact the performance of various quantum algorithms for the JOO problem.
\item a runtime complexity analysis that shows that the runtime of our approach is $O(2^m-m)$ with $m$ number of relations, which is theoretically optimal for exact methods using direct join costs (i.e., join costs for each possible join). 
\end{itemize}

The remainder of this paper is organized as follows: Section \ref{sec:basics} provides an introduction to quantum computing fundamentals, various quantum algorithms, and key concepts related to QUBO. It also covers query optimization and a review of related work. Section \ref{sec:concept} presents the QUBO formulation of the join ordering problem, details the proposed method of Eliminating Cartesian Product (ECP) and selectivity along with SQSS, and includes a comparative runtime complexity analysis. Section \ref{sec:experiments} conducts an extensive experimental evaluation and provides a detailed analysis of the evaluation results. Finally, section \ref{sec:conclusions} summarizes our findings and presents concluding remarks.

\section{Basics}
\label{sec:basics}

In this section, we introduce the fundamental concepts of QUBO, which is used to formulate the join-ordering problem.
Additionally, we provide an overview of quantum computing and various quantum algorithms utilized for quantum optimization, exploring their potential applications in the database management systems (DBMS) community.
Furthermore, we discuss the principles of query optimization and review relevant related work.

\subsection{Quadratic Unconstrained Binary Optimization}
\label{sec:intro_qubo}

Quadratic Unconstrained Binary Optimization (QUBO) is commonly used for minimizing a quadratic function (Hamiltonian) over $N$ binary variables with $2^N$ possible states. 
The quadratic function to be minimized is called the cost function or energy, and it can be written as:
\begin{equation}
    E_{qubo}(x) = \alpha + \sum_i \alpha_i x_i + \sum_{\langle{i_1,i_2}\rangle}\alpha_{i_1,i_2}x_{i_1} x_{i_2}
\end{equation}
where $x = (x_1,x_2,...,x_N)$ represents the assignments of $N$ binary variables and $\langle{i_1,i_2}\rangle$ indicates a pair of binary variables, those with indices $i_1,i_2$.  
Binary variables have values $x_i \in \{0,1\}$, and coefficients $\alpha, \alpha_i, \alpha_{i_1,i_2}$ are real.

QUBO models are versatile and widely applicable to problems in areas including allocating resources, clustering, set partitioning, locating facilities, solving assignment problems, and solving sequencing problems \cite{kochenberger2014unconstrained, glover2022quantum}.
The Ising spin model with two-body interactions, which is a physical analogue to QUBO, establishes a bridge between computational optimization and statistical physics \cite{baxter109rodney}.
Many NP-hard problems can be directly formulated as QUBOs, making them pivotal in the study of computational complexity \cite{feige1995approximating, lucas2014ising}.

QUBO is being widely used to implement the optimization problem on variational quantum algorithms for tackling large-scale problems.
For instance, the integration of hybrid quantum-classical methods, such as VQEs and QAOAs, offers promising approaches to solving QUBOs with better scalability and precision \cite{farhi2014quantum, cerezo2021variational, nayak2024quantum}.
In addition, modern developments have explored more efficient embedding techniques for mapping QUBOs onto hardware, such as D-Wave’s quantum annealers, which exploit the quantum annealing algorithm for optimization and enhancing the performance through preconditioning and constraint reduction \cite{glover2022quantum, nayak2024quantum}.

\subsection{Fundamentals of Quantum Computing}

Quantum computing is a field of computing that harnesses the principles of quantum mechanics, like superposition, entanglement, and quantum interference, to perform calculations that classical computers cannot efficiently solve.
The building block of quantum computing is the qubit, which is the quantum analogue of the classical bit, with two orthonormal basis states $|0\rangle$ and $|1\rangle$.
A qubit state can be defined as:
\begin{equation}
    |\psi\rangle =\alpha|0\rangle + \beta|1\rangle
\end{equation}
Where $\alpha$ and $\beta$ are complex numbers, which satisfy the condition; $|\alpha|^2 + |\beta|^2 = 1$.

Quantum gates are basically used to perform operations on qubits, which manipulate the qubit.
Quantum gates can be represented by a unitary matrix, ensuring the preservation of the quantum state normalization.
There are several quantum gates, like Pauli gates, Hadamard gates, and the CNOT gate.

\subsection{Quantum Circuit Models}

Quantum circuit models are the cornerstone of quantum computation, providing a framework for representing quantum algorithms using a sequence of quantum gates applied to qubits.
These models are analogous to classical logic circuits, which use the principles of quantum mechanics like superposition and entanglement, and utilize unitary transformations performed by quantum gates to manipulate quantum states efficiently, which distinguishes quantum computing from classical approaches \cite{nielsen2010quantum, arute2019quantum}.
Recently, these models have seen the integration of machine learning techniques to enhance circuit synthesis, e.g., diffusion models \cite{furrutter2024quantum}.

Furthermore, it serves as the foundation for hybrid approaches like VQC, which combine classical optimization with quantum resources for solving optimization problems \cite{cerezo2021variational}.
Quantum circuit models are compatible with current hardware architectures, including superconducting qubits and trapped ions, and are widely used in VQAs, quantum error correction, and fault-tolerant computing \cite{preskill2018quantum}.
Although to improve computational efficiency, new quantum circuit architectures are being proposed, e.g., a diamond-shaped quantum circuit has been designed to approximate multi-qubit gate-based circuits, and it offers significant advantages over traditional constructions, particularly in handling highly entangled quantum states \cite{miyakoshi2023diamond}.
We will now discuss quantum algorithms that we used for quantum optimization to find the ground state of the QUBO problem.

\subsubsection{Variational Quantum Eigensolver}

Variational Quantum Eigensolver (VQE) is a hybrid quantum algorithm that is developed to estimate the ground-state energy of a given quantum system \cite{peruzzo2014variational}. For this purpose, an approximate Hamiltonian of a physical system is designed using a linear combination of the tensor products of Pauli operators, as in Equation~\ref{eq:VQEHamiltonian}.

\begin{equation}
    H\approx\hat{H}=\sum_i c_i P_i
    \label{eq:VQEHamiltonian}
\end{equation}
Here, each $c_i$ is a constant coefficient and each $P_i$ is a tensor product of $n$ Pauli operators (one of $I$, $\sigma^X$, $\sigma^Y$ or $\sigma^Z$). With this formulation, the expected value of each operator in the sum, $\left\langle P_i \right\rangle$, can be computed separately for a given quantum state. Then the expected value for the total Hamiltonian can be calculated as

\begin{equation}
    \left\langle \hat{H} \right\rangle = \sum_i c_i\left\langle P_i \right\rangle
\end{equation}
The purpose of this simplification is to reduce the coherence time of each quantum calculation, which makes VQE applicable for near-term quantum hardware.

Since the aim of the algorithm is to find the energy of the ground state, during the course of VQE, $\left\langle \hat{H} \right\rangle$ is calculated for many different quantum states. The search space of states is represented by a variational quantum circuit, which consists of parameterized single-qubit rotation operators, and two-qubit entanglement gates. The structure of the ansatz is selected to work efficiently on NISQ-era hardware, while also being able to represent a large set of possible eigenstates.

Despite its initial intention of calculating the ground state energy of molecules, as a natural step, the ability to find the state with minimum energy is adapted to mathematical optimization. VQE is applied to many optimization problems that can be modelled as a QUBO formulation.

\subsubsection{Quantum Approximate Optimization Algorithm}
Quantum approximate optimization algorithm (QAOA) is developed from the inspiration of the adiabatic principle of quantum mechanics similar to QA \cite{farhi2014quantum}. It can be said to be a discrete approximation of QA designed for gate-based quantum computers.

In the beginning, the initial quantum state is set to the equal superposition of all possible solutions.

\begin{equation}
    \left | s \right\rangle = \sum_{j=1}^{2^n} = \left | j \right\rangle
\end{equation}
Then, this state is evolved by applying unitary operations for $p$ times where $p\geq1$. Each evolution operator evolves the state with respect to the \textit{cost Hamiltonian} $H_C$ and a \textit{mixer Hamiltonian} $H_M$ and the amount of evolution depends on 2 parameters for each unitary operator. The $k^{th}$ unitary operator is defined as

\begin{equation}
    U_k(\gamma_k,\beta_k)=e^{-i\beta_k H_M}e^{-i\gamma_k H_C}
\end{equation}
where the structure of $H_C$ depends on the formulation of the problem (most probably a QUBO formulation) and $H_M$ is defined as

\begin{equation}
    H_M=\sum_{j=1}^n \sigma_j^X .
\end{equation}
Then, the final state after applying the unitary operations is

\begin{equation}
    U_p(\gamma_p,\beta_p)U_{p-1}(\gamma_{p-1},\beta_{p-1})...U_1(\gamma_1,\beta_1)\left | s \right\rangle
\end{equation}

During the course of QAOA, a quantum circuit is run many times to measure the final state using different $\gamma_k$ and $\beta_k$ parameters. The $2p$ parameters are optimized using a classical computer to get a more desirable cost value. This algorithm can be applicable to any mathematical optimization problem that can be written in a QUBO formulation, some of which are given in section \ref{sec:intro_qubo}.

\subsection{Quantum Annealing}
The motivation for quantum annealing (QA) lies in the adiabatic theorem \cite{morita2008mathematical}, which states that a quantum system starting in its ground state will remain there if the Hamiltonian governing it changes slowly enough.
It utilizes quantum fluctuations to search for the ground state of a problem Hamiltonian.
The rate of change is limited by the smallest energy gap between the ground state and the first excited state during evolution.
QA is started by preparing the system in the ground state of an initial Hamiltonian, which is known and easy to prepare, and is denoted as $H_i$.
Then, the system is slowly changed so that the contribution of $H_i$ is slowly reduced while the magnitude of a final (also known as the target) Hamiltonian, denoted as $H_f$, is increased using the time parameter $t$.
\begin{equation}
    H(t) = A(t)*H_i + B(t)*H_f
\end{equation}
where $t \in [0,T_a]$, the annealing schedule is defined by functions $A(t)$ and $B(t)$ with $A(0)=1, B(0)=0$ at initial state and $A(T_a)=0, B(T_a)=1$ at final state so that H interpolates between $H_i$ at $t_i$ and $H_p$ at $t_f$.

In the current state-of-the-art, D-Wave QPU utilizes quantum annealing to perform quantum optimization, and these quantum annealers are programmable hardware implementations of quantum annealing that use superconducting flux qubits \cite{lanting2014entanglement,king2022coherent}.
In recent times, quantum annealers have more than $5000$ qubits, such as the D-Wave Advantage\_system4.1.

\subsection{Query Optimization}
\label{sec:intro_QO}

Query optimization is a crucial process in DBMS that enhances query efficiency while ensuring accuracy and consistency. 
The objective is to minimize execution time and resource consumption by selecting the most efficient execution plan \cite{ramakrishnan2003database, silberschatz2011database}. 
Traditional cost-based optimization evaluates multiple plans based on CPU utilization, disk I/O, and memory requirements, selecting the one with the lowest estimated cost \cite{jarke1984query}.

As database workloads evolve, modern approaches incorporate machine learning, such as learned cost models and reinforcement learning, to refine execution plan selection \cite{marcus2019neo}. 
Adaptive query processing techniques, such as adaptive recursive query optimization, enable real-time adjustments to enhance performance in complex scenarios \cite{herlihy2024adaptive}.

Recent research has further integrated AI-driven methods, including artificial bee colony algorithms \cite{du2024query} and predictive modeling \cite{rahman2024advanced}, demonstrating significant improvements in distributed databases. 
These advancements shift query optimization from static, rule-based methods to intelligent, self-adaptive strategies for large-scale data environments. 
In this paper, we focus on join order optimization as a critical aspect of query optimization.

\subsubsection{Join Order Optimization}
\label{sec:intro_joo}

Join-order optimization is a crucial challenge in relational database query processing, as it significantly enhances query performance in large-scale data processing systems. By optimizing the join sequence, databases can achieve faster analytics, enable real-time decision-making, and minimize computational costs \cite{cabrera2023efficiently, leis2018query}.

Given the factorial growth in the number of possible join orders, the problem becomes combinatorial in nature. It can be reformulated as a Quadratic Unconstrained Binary Optimization (QUBO) problem, allowing the application of advanced optimization techniques such as quantum computing and heuristic algorithms \cite{khan2023quantum, cabrera2023efficiently}.

Traditionally, cost-based optimization techniques estimate the execution cost of different join sequences by considering factors such as data statistics, cardinality, and available indexes. These cost models are heavily based on accurate cardinality estimation, typically derived from real-world databases \cite{han2022cardinality, leis2018query}. However, in practical scenarios, cardinality estimation often incorporates simplifying assumptions, such as uniform data distribution and attribute independence, to reduce computational complexity \cite{chaudhuri2023analyzing}.

Unfortunately, these assumptions frequently fail to hold in real-world datasets, leading to inaccurate estimates and, consequently, suboptimal or even highly inefficient query execution plans. Research by \cite{leis2018query} highlights the critical impact of cardinality misestimation, demonstrating how erroneous assumptions can degrade query performance substantially. 
Furthermore, recent work \cite{ebergen2022join} investigates scenarios in which optimizers function with limited statistical information, further emphasizing the consequences of inaccurate cardinality estimates in execution plans.

\subsection{Further Related Work}
To investigate the join order optimization, algorithms such as dynamic programming \cite{selinger1979access}, heuristic algorithms, randomized algorithms, genetic algorithms \cite{steinbrunn1997heuristic}, the ant colony optimization algorithm \cite{li2008application}, particle swarm optimization \cite{mingyao2015embedded}, and machine learning \cite{marcus2018deep} have been implemented.

There are quantum algorithms, which have been utilized to investigate the join-order optimization problem using quantum processing units like quantum annealers, gate-based quantum computers, and quantum simulators. Methods like mixed-integer linear programming \cite{schonberger2023ready}, variational quantum eigensolvers, quantum approximation optimization algorithms, quantum annealing, simulated annealing, and quantum machine learning have been tested while exploiting variational quantum circuits \cite{winker2023quantum}.

In the work \cite{schonberger2023ready}, the left/right deep join order has been investigated using quantum annealing by converting the join order problem to the QUBO problem.
In the paper \cite{schonberger2022quantum}, the authors have proposed a theoretical and mathematical approach to solve the join-order problem by converting it to the QUBO model.
It has been claimed that the QUBO problem for join-ordering can be solved for general bushy trees, but no experimental data have been provided for the approach.

In the paper \cite{nayak2023constructing}, the author has solved the join order problem for joining three, four, and five relations.
It converted the join order problem to the QUBO problem, supporting bushy trees as well in the solutions set (including left/right-deep join trees).

In the paper \cite{nayak2024quantum}, the author has proposed a hybrid approach by partitioning the QUBO search space to be solved by the D-Wave quantum computer and universal quantum simulator. 
It can find the optimal join order of up to 7 relations on a quantum annealer, as well as simulated annealing. 
It solved artificial queries for up to 5 relations using the universal quantum simulator.

For our proposed novel method of ECP-SQSS, we build upon the work in \cite{nayak2024quantum} to study the impact of incorporating the Eliminating the Cartesian Product technique with the Splitting the QUBO Search Space on join order optimization.
The proposed approach has been evaluated using various algorithms, including QA, SA, QAOA, and VQE. 
Specifically, we conducted experiments with QA on D-Wave’s QPU and evaluated the QAOA and VQE algorithms on a universal quantum simulator by Qiskit.
Additionally, we further analyzed the effect of selectivity on the performance of D-Wave’s QPU for join order optimization.

\section{Join Order Optimization improved with ECP and Selectivity}
\label{sec:concept}

In this section, we define the join-ordering problem in DBMS. We introduce techniques such as ECP and selectivity, and how join ordering can be modeled as a QUBO problem, which can be solved using quantum algorithms. The QUBO formulation represents the join conditions, costs, and constraints as a binary optimization problem. By leveraging QUBO-based optimization, we encode the problem into a target problem suitable for quantum algorithms, facilitating efficient query execution. ECP and selectivity make the formulation more scalable and suitable for quantum hardware.

\subsection{Defining Join Ordering Problem}
\label{sec:def}

A join operation in a relational query combines two relations, and for queries involving more than two relations, the process includes intermediate results. 
At each step, two relations (or intermediate results) are joined until a single final relation remains, representing the fully joined query result. 
For simplicity, we assume that the join operation is commutative, which means that $R_1 \bowtie R_2$ is equivalent to $R_2 \bowtie R_1$. 
Consequently, the join operations can be structured as a binary tree, where:
\begin{itemize}
    \item Leaf nodes represent individual relations.
    \item Internal nodes denote intermediate join results.
    \item The root contains the final result of the join.
\end{itemize}
This structure is referred to as a join tree. 
Each node in the tree is labelled with the set of relations already joined at that point, and the binary tree is completely defined by the sets of all its inner nodes.
For illustration purposes, check Figure~\ref{fig:ExampleValidJoinTree}.

For a given query with a set of relations:$M:= \{R_1, R_2, \dots, R_m\}$,
each node in the join tree represents a subset of \(M\), and the complete join tree is a subset of the power set \(P\), which contains all possible subsets of \(M\). 
Our objective is to determine the join tree with the lowest computational cost. 
However, not all subsets of the power set represent valid join orders (see Section~\ref{sec:JOO_as_QUBO}).
For example, the power set for three relations \(R_1\), \(R_2\), and \(R_3\) is:
\begin{align*}
    P =  \{& \{\}, \{R_1\}, \{R_2\}, \{R_3\}, \{R_1, R_2\}, \{R_1, R_3\},\\
        & \{R_2, R_3\}, \{R_1, R_2, R_3\} \}
\end{align*}
A join order such as \( (R_1 \bowtie R_2) \bowtie R_3 \) can be represented as:
\[
\{\{R_1\}, \{R_2\}, \{R_3\}, \{R_1, R_2\}, \{R_1, R_2, R_3\} \}.
\]
If it is required to eliminate Cartesian products in the final join tree, then to reduce complexity, we eliminate redundant sets:
\begin{itemize}
    \item The empty set is unnecessary in this representation.
    \item Single relations and the final result are always included, so they do not need to be explicitly stored.
    \item Eliminating Cartesian product
\end{itemize}
In the above case, if the cross-join between the relations, $\{R_1, R_3\}$, then the reduced power set for three relations is:
\[
P_{ecp} = \{\{R_1, R_2\}, \{R_2, R_3\}\}.
\]
The join order \( (R_1 \bowtie R_2) \bowtie R_3 \) can now be modeled as: $\{\{R_1, R_2\}\}$.
To formulate this as a binary optimization problem, we define a binary variable for each element in the reduced power set, where each variable indicates whether the corresponding subset is included in the join-order solution. 
Thus, we transform the join-ordering problem into a QUBO problem, making it suitable for optimization using quantum and classical methods.

\subsection{Eliminating the Cartesian Product (ECP)}
\label{sec:ecp_theory}
The Cartesian product (also called cross-join) returns all combinations of rows from the two relations of the join: each row in the first relation is paired with all the rows in the second relation.
Cross-join can be denoted as:
\begin{equation}
    R_i \times R_j = \{ (a, b) \ | \ a \in R_i, \ b \in R_j \}
\end{equation}
Where $R_i,R_j\in M$, $a$ is an element (row) from table $R_i$, $b$ is an element (row) from table $R_j$ and $R_i \times R_j$ is the set of all possible ordered pairs $(a,b)$.

Cartesian products should be avoided in join trees as they produce a large number of rows in the result set in database queries that conduct cross-join operations.
This may have a negative effect on the query's performance and cause it to execute more slowly.
To avoid Cartesian products, it is essential to properly define the join conditions between the relations, or a pair of relations should share a foreign key relationship.
Intermediate subsets of relations that involve Cartesian Products should be eliminated. Mathematically, it can be understood. 
Let's suppose for a given query $Q$ relations $R_i$ and $R_j$ are joinable if $Q$ has a join condition between attributes of the relations $R_i$ and $R_j$, then $J_R$ can be defined as:
\begin{equation}
    J_R=\{(R_i,R_j)|R_i,R_j\in R \wedge \mbox{$R_i$ and $R_j$ are joinable}\}
\end{equation}
where $R\subseteq \{R_1,...,R_m\}$.
The reduced power set $P_{ecp}$, after eliminating the cartesian products, is defined as:
\begin{equation}
\begin{split}
    P_{ecp} & =\{ p | p\in P \setminus \{R_1,...,R_m\} \wedge|p|\geq 2 \\
    & \wedge \forall R_i \in p: \forall R_j \in p\setminus \{R_i\}: (R_i,R_j)\in J_p^+\}
\end{split}
\end{equation}
where $J_R^+$ is the transitive closure of $J_R$.
By specifying the appropriate join conditions, you can limit the number of rows that are combined, resulting in a more efficient query.
The performance of your database queries can be improved by avoiding Cartesian products and optimizing the join conditions.

\subsection{Enhancing Database Efficiency: ECP use cases}
\label{sec:use_cases_ecp}

ECP is crucial for improving database performance, optimizing query execution, and ensuring efficient resource utilization \cite{morishita1997avoiding}. Cartesian products generate a large number of intermediate rows, significantly slowing down query execution. Avoiding Cartesian products reduces the computational complexity, particularly in queries involving large tables. It generates a large number of intermediate rows, which consumes substantial memory \cite{selinger1979access}, and modern query optimizers attempt to prune unnecessary Cartesian products to avoid excessive memory allocation. 

In large-scale Extract, Transform, Load (ETL) operations, Cartesian products can lead to unnecessary joins, increasing data processing time. 
Eliminating them ensures ETL pipelines process only meaningful data, reducing execution time \cite{fiebig2002anatomy}. Query optimizers in SQL engines use heuristics and cost-based approaches to minimize Cartesian products \cite{silberschatz2011database}. Proper indexing and join ordering prevent the optimizer from choosing suboptimal execution plans. In distributed database systems \cite{graefe1993query}, a Cartesian product can lead to excessive data transfer between nodes. Eliminating it ensures that only necessary data is transmitted, reducing network latency and bandwidth consumption. Cartesian products in SQL-based machine learning workflows can increase training times by introducing redundant data \cite{zaharia2016apache}. Therefore, eliminating cartesian products ensures that feature engineering and aggregation operations are performed efficiently.

\subsection{Utilising ECP for Join-Ordering}
\label{sec:Utilisation_of_ECP_for_JOO}
We present a straightforward method of ECP. 
When formulating join ordering as QUBO, ECP aids in reducing the binary variables required for it, and consequently, the constraints.
For example, for a given set of relations $M:=\{R_1, R_2, R_3, R_4\}$, the power set ($P$) can be defined before eliminating the cartesian product as: 
\begin{align*}
P = \{ &  \{R_1, R_2\}, \{R_1, R_3\}, \{R_1, R_4\}, \{R_2, R_3\},\\
       &  \{R_2, R_4\}, \{R_3, R_4\}, \{R_1, R_2, R_3\}, \\
       & \{R_1, R_2, R_4\}, \{R_1, R_3, R_4\}, \{R_2, R_3, R_4\},\\
       & \{R_1, R_2, R_3, R_4\}\}.
\end{align*}
Let's suppose there are relations which are not joinable $(R_1,R_3),\\
(R_1,R_4), (R_2,R_4)$ therefore
\begin{align*}
    J_{\{R_1, R_2, R_3, R_4\}} = \{ & (R_1,R_2), (R_2,R_1), (R_2,R_3), \\
    &  (R_3,R_2), (R_3,R_4), (R_4,R_3)\}
\end{align*}
and
\begin{align*}
J_{\{R_1, R_2, R_3, R_4\}}^+ = \{ & (R_1, R_3), (R_3, R_1), (R_1, R_4),\\
& (R_4, R_1), (R_2, R_3), (R_3, R_2),\\
&  (R_2,R_4), (R_4, R_2), (R_3,R_4)\\
&  (R_4, R_3)\}
\end{align*}

After eliminating the cartesian products, the power set will have the binary variables with a join condition, and then the reduced power set becomes:
\begin{align*}
P_{ecp} = \{ & \{R_1, R_4\}, \{R_2, R_3\}, \{R_3, R_4\}, \\
       & \{R_1, R_3, R_4\}, \{R_2, R_3, R_4\}\}.
\end{align*}
Now we have a reduced power set $P_{ecp}$ of intermediate joins, which reduces the constraints in the QUBO formulation and uses fewer qubits to run the experiments on hardware and simulator.

\subsection{Selectivity}
\label{intro_selectivity}
In database management systems, selectivity in SQL queries measures the fraction of rows in a database table that satisfy a given predicate.
Where a predicate refers to a condition or expression that evaluates to either true, false, or sometimes unknown (in cases of null values). 
Predicates are used in SQL queries to filter data by specifying the criteria that rows must satisfy to be included in the query result.
Predicates are most commonly used in clauses such as:
\begin{lstlisting}[language = SQL,
           showspaces=false,
           basicstyle=\ttfamily,
           commentstyle=\color{gray}]
SELECT COUNT(*) FROM 
constructorresults,
WHERE 
constructorresults.raceId > k;
\end{lstlisting}

A predicate in the above SQL query is:

\begin{lstlisting}[language = SQL,
           showspaces=false,
           basicstyle=\ttfamily,
           commentstyle=\color{gray}]
constructorresults.raceId > k
\end{lstlisting}

Let’s $R$ be a relation (table) with  $|R|$ rows (cardinality of $R$).
$p$ be a predicate (filter condition) applied to $R$.
$R_p$  be the subset of rows in $R$ that satisfy the predicate $p$.
It is calculated as the ratio of the number of rows that match the predicate to the total number of rows in the table \cite{lynch1988selectivity}.
\begin{equation}
    \text{Selectivity}(p, R) = \frac{|R_p|}{|R|}
\end{equation}
where $|R_p|$ number of rows satisfies the predicate and $|R|$ is the total number of rows.

\subsection{Using Selectivity to modify assigned weights to binary variables}
\label{sec:using_selectivity}

On performing selection on the dataset, selectivity effectively reduces the size of the data by decreasing the number of rows (cardinality) that satisfy the selection condition.
The weight ($w_i$) assigned to a binary variable in the QUBO formulation is proportional to the cardinality of the data satisfying all the join conditions.
We use predicted cardinalities as the weights for the QUBO formulation. 
The cardinality for the join of a set of relations is independent of the order in which they are joined. 
We can use existing cardinality estimators by choosing a random join tree and predicting the cardinality. 
We do this for every $R \in P$.

With changes in cardinality due to selectivity, the weight distribution of the binary variables associated with the join conditions also changes.
This variation in the weight distribution modifies the QUBO formulation, which impacts the energy landscape. 
The modified energy landscape may simplify the problem, making it easier for quantum algorithms to locate the global minimum.
To illustrate the change in weight distribution using selectivity, consider a binary variable representing the join between two relations, $R_1$ and $R_2$, with some initial cardinality. 
When a selection condition is applied to $R_1$, reducing its number of rows, the cardinality of the join between $R_1$ and $R_2$ decreases, altering the associated weight in the QUBO formulation.
This change in cardinality directly influences the energy landscape of the QUBO problem as it changes the weight of binary variables in the QUBO.
By adjusting the dataset’s cardinality, selectivity modifies the QUBO problem in a way that can lead to a smoother energy landscape.
This can make it easier for quantum algorithms to converge to the global minimum.

In our work, where selectivity is applied to real-world queries combined with the SQSS method, we adopted the QUBO formulation detailed in \cite{nayak2024quantum}.
In this previous work, the SQSS method was thoroughly described, including all the necessary constraints to be considered when implementing it.

\subsection{Join-Ordering as QUBO problem}
\label{sec:JOO_as_QUBO}

The power set $P$ represents joins with associated processing costs $(w_i)$, where each $w_i$ corresponds to a single join for $i \in P$. 
Binary variables $x_i$ indicate whether a specific join in $P$ is part of the overall join to be processed. 
The QUBO formulation can be divided into subtotals, each representing smaller, self-contained problems.
To prioritize joins with lower costs, a constant $w_{\text{max}}$ is subtracted from each $w_i$.
The constant is defined as:
\begin{equation}
\label{eq:max_weight}
    w_{\text{max}} = \max(w_i) + c 
\end{equation}
where $c > 0$, ensures $w_{\text{max}}$ is larger than the highest weight in the current instance. 
Subtraction of $w_{\text{max}}$ shifts' lower weights into the negative range, favoring them during optimization.  

The first constraint of the QUBO formula is expressed as follows:
\[
C_1(P) = \sum_{x \in P} x \cdot (w_i - w_{\text{max}})
\]
This power set can be reduced further as we discussed in section \ref{sec:def}, and, for our new method of ECP, it further reduces the possible cross-joins condition in the power set by elimination of the cartesian product; the power set becomes $P_{ecp}$.

With this, the first constraint modifies as
\begin{equation}
    C_1(P_{ecp}) = \sum_{x \in P_{ecp}} x \cdot (w_i - w_{\text{max}})
\end{equation}
Here, $C_1$ minimizes the sum, where all weights are negative. 
Without constraints, all binary variables would be set to 1, resulting in an invalid solution containing all possible joins.
To ensure valid join combinations, constraints are introduced. 
These constraints penalize invalid join trees by applying $w_{\text{max}}$ as a penalty weight to their corresponding binary variables, ensuring the solution forms a valid join tree.

\begin{figure}[htb!]
     \centering
     \begin{subfigure}[b]{0.3\textwidth}
         \centering
\begin{tikzpicture}
 \node {$\mbox{\Huge$\bowtie$}_{\{R_1,R_2,R_3,R_4,R_5\}}$} [sibling distance = 2.4cm, level distance = 1cm]
    child {node {$\mbox{\Huge$\bowtie$}_{\{R_1,R_2,R_3\}}$} [sibling distance = 1.2cm]
        child {node {$R_1$}}
        child {node {$\mbox{\Huge$\bowtie$}_{\{R_2,R_3\}}$}
            child {node {$R_2$}}
            child {node {$R_3$}}
            }
    } 
    child {node {$\mbox{\Huge$\bowtie$}_{\{R_4,R_5\}}$} [sibling distance = 1.2cm]
        child {node {$R_4$}}
        child {node {$R_5$}}
    };
 \end{tikzpicture}
        \caption{Example}
        \label{fig:ExampleValidJoinTree}
     \end{subfigure}
     \hfill
     \begin{subfigure}[b]{0.15\textwidth}
         \centering
\begin{tikzpicture}
\node {...} [sibling distance = 1.2cm, level distance = 0.7cm]
  child {node {$\mbox{\Huge$\bowtie_{z}$}$} 
    child {node {} [sibling distance = 1.2cm]
        child {node{...}
            edge from parent [solid]
        }
        child {node {$\mbox{\Huge$\bowtie_{y}$}$}
            child {node{...}
                edge from parent [solid]
            }
            child {node{...}
                edge from parent [solid]
            }
            edge from parent [dotted]
        }
        edge from parent [solid]
    } 
    child {node {...}
        edge from parent [solid]
    }
    edge from parent [dotted]
  };
 \end{tikzpicture}
 {\Large$y \subseteq z$}
         \caption{Constraint}
         \label{fig:ConstraintValidJoinTree}
     \end{subfigure}
        \caption{Example and constraint for valid join trees}
        \label{fig:ValidJoinTree}
\end{figure}
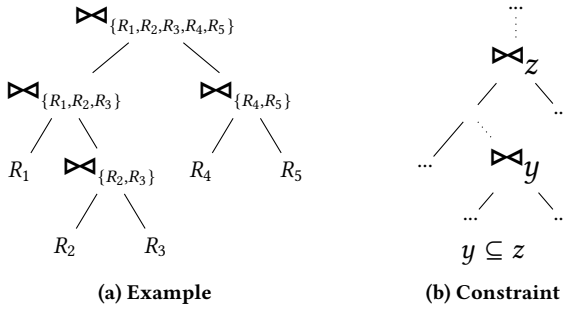
In the illustration of valid join trees (see Figure~\ref{fig:ValidJoinTree}), the set $z$ of relations in a join $\bowtie_z$ always includes the relations $y$ of any join $\bowtie_y$ that appears at any depth below $\bowtie_z$ (see Figure~\ref{fig:ConstraintValidJoinTree}). 
Additionally, a relation $R_i$ appears exactly once in a valid join tree as a leaf node, and all relations in $z$ must be leaf nodes in the subtree of $\bowtie_z$. 
Any other combinations violate the rules of a valid join tree and must be penalized.
If sets $y$ and $z$ share relations but one is not a subset of the other, any solution including both is invalid and requires a penalty. The set of invalid joins caused by $y$ is defined as:
\[
W_{ecp}(y) := \left\{z \mid z \in P_{ecp} \wedge |z| \geq |y| \wedge z \cap y \neq \emptyset \wedge y \nsubseteq z\right\}
\]
Using $W_{ecp}(y)$, we define the penalty term $C_2$ for invalid combinations in the QUBO formulation:
\begin{equation}
    C_2(P_{ecp}) = \sum_{y \in P_{ecp}} \sum_{z \in W_{ecp}(y)} x_y \cdot x_z \cdot w_{\text{max}}
\end{equation}
The global minima of $C_2$ represent all valid join trees.
To find the optimal valid join tree with the lowest cost, we minimize the combined objective function:
\begin{equation}
\label{eq:final_qubo_joo}
    C(P_{ecp}) = C_1(P_{ecp}) + C_2(P_{ecp})
\end{equation}

\subsection{Worked example for ECP as QUBO problem}

In this section, we demonstrate the QUBO formulation by using an example for optimal join orders with and without using the ECP method to solve the join-ordering problem.
We will continue with the same example discussed in the section \ref{sec:Utilisation_of_ECP_for_JOO}.
First, we solve the QUBO without eliminating the cartesian product where we had a complete power set:
\begin{align*}
P = \{ & \{R_1, R_2\}, \{R_1, R_3\}, \{R_1, R_4\},\\
             & \{R_2, R_3\}, \{R_2, R_4\}, \{\{R_3, R_4\},\\
             & \{R_1, R_2, R_3\}, \{R_1, R_2, R_4\}, \{R_1, R_3, R_4\},\\
             & \{R_2, R_3, R_4\}, \{R_1, R_2, R_3, R_4\}\}.
\end{align*}
Binary variables represent each subset $S \in P$, indicating whether the relations in $S$ are joined. 
These variables, however, do not specify the exact join sequence but ensure the relations in $S$ are part of the final join result.
For instance, it $\{R_1, R_2, R_3\}$ indicates the join of $R_1$, $R_2$, and $R_3$, without specifying the sequence. 
If it $\{R_1, R_3\}$ is also included, the join order becomes $(R_1 \bowtie R_3) \bowtie R_2$, and we assign the weights to each binary variable; these weights are based on join output size and exclude costs for input relations or subjoins.
To begin with, let's suppose the assigned weights to the binary variables with their representation are as $x_i$ in the QUBO formulation as
\begin{align*}
    & w_{\{R_1, R_2\}} = x_0 = 3, \quad w_{\{R_1, R_3\}} = x_1 = 9,\\
    & w_{\{R_1, R_4\}} = x_2 = 4, \quad w_{\{R_2, R_3\}} = x_3 = 6,\\
    & w_{\{R_2, R_4\}} = x_4 = 5, \quad w_{\{R_3, R_4\}} = x_5 = 1\\
    & w_{\{R_1, R_2, R_3\}} = x_6 = 10,\quad w_{\{R_1, R_2, R_4\}} = x_7 = 3,\\
    & w_{\{R_1, R_3, R_4\}} = x_8 = 7,\quad w_{\{R_2, R_3, R_4\}} = x_9 = 8,\\
    & \text{and } \quad w_{\{R_1, R_2, R_3, R_4\}} = x_{10} = 2.
\end{align*}
Maximum weight can be calculated according to the equation \ref{eq:max_weight}:
\[
w_{\text{max}} = \max(\text{weights}) + 1 = 10 + 1 = 11
\]
The target function C(x) is formulated according to the equation \ref{eq:final_qubo_joo} as:
\begin{align*}
    C(x) = & \ 11 \cdot x_0 x_1 + 11 \cdot x_0 x_2 + 11 \cdot x_1 x_2 + 11 \cdot x_0 x_3 \\
           & + 11 \cdot x_1 x_3 + 11 \cdot x_0 x_4 + 11 \cdot x_2 x_4 + 11 \cdot x_3 x_4 \\
           & + 11 \cdot x_1 x_5 + 11 \cdot x_2 x_5 + 11 \cdot x_3 x_5 + 11 \cdot x_4 x_5 \\
           & + 11 \cdot x_2 x_6 + 11 \cdot x_4 x_6 + 11 \cdot x_5 x_6 + 11 \cdot x_1 x_7 \\
           & + 11 \cdot x_3 x_7 + 11 \cdot x_5 x_7 + 11 \cdot x_6 x_7 + 11 \cdot x_0 x_8 \\
           & + 11 \cdot x_3 x_8 + 11 \cdot x_4 x_8 + 11 \cdot x_6 x_8 + 11 \cdot x_7 x_8 \\
           & + 11 \cdot x_0 x_9 + 11 \cdot x_1 x_9 + 11 \cdot x_2 x_9 + 11 \cdot x_6 x_9 \\
           & + 11 \cdot x_7 x_9 + 11 \cdot x_8 x_9 - 8 \cdot x_0 - 2 \cdot x_1 - 7 \cdot x_2\\
           & - 5 \cdot x_3 - 6 \cdot x_4 - 10 \cdot x_5 - x_6 - 8 \cdot x_7 - 4 \cdot x_8\\
           & - 3 \cdot x_9 - 9 \cdot x_{10}
\end{align*}
Linear terms penalize high-cost joins, while the quadratic term penalizes invalid join combinations.
For example, it $x_0 \cdot x_1$ imposes a penalty for the combination:$x_0x_1 \rightarrow ((R_1 \bowtie R_2) \bowtie (R_1 \Join R_3))$.
On solving for the QUBO problem, we get the minimum value of the function $C(x)$:

$\text{fval} = -27$ with $x_0 = 1, \quad x_5 = 1, \quad x_{10} = 1$\\
\noindent
Therefore, binary variables in the solution set is: $\{\{R_1, R_2\}, \{\{R_3, R_4\},\\
\{R_1, R_2, R_3, R_4\}\}$ and the optimal join order for the above problem is given as
\[
(R_1 \bowtie R_2) \bowtie (R_3 \bowtie R_4)
\]
\noindent
For the given example in section \ref{sec:Utilisation_of_ECP_for_JOO} with relations that are not joinable, $(R_1, R_3), (R_1, R_4), (R_2, R_4)$ as we discussed, the valid intermediate joins are contained in the reduced power set:
\begin{align*}
P_{ecp} = \{ & \{R_1, R_2\}, \{R_2, R_3\}, \{R_3, R_4\}, \{R_1, R_2, R_3\},\\
       &  \{R_1, R_2, R_4\}, \{R_2, R_3, R_4\}\}.
\end{align*}
The maximum weight is calculated as:
\[
w_{\text{max}} = \max(\text{weights}) + 1 = 10 + 1 = 11
\]
The target function for the above QUBO problem will look like this:
\begin{align*}
    C(x) = & \ 11 \cdot x_0 x_3 + 11 \cdot x_3 x_5 + 11 \cdot x_5 x_6 + 11 \cdot x_3 x_7 \\
           & + 11 \cdot x_5 x_7 + 11 \cdot x_6 x_7 + 11 \cdot x_0 x_9 - 8 \cdot x_0 \\
           & - 5 \cdot x_3 - 10 \cdot x_5 - x_6 - 8 \cdot x_7 - 3 \cdot x_9
\end{align*}
On solving this QUBO problem, which had reduced quadratic and linear terms using the ECP method, we get the minimum value of the target function and binary variable with value $1$:

$\text{fval} = -18$ with $x_0 = 1, \quad x_5 = 1$\\
\noindent
The solution set contains the joins: $\{R_1, R_2\}$ and $\{R_3, R_4\}$. 
This corresponds to the join order $(R_1 \bowtie R_2) \bowtie (R_3 \bowtie R_4)$,  to which the last variable can finally join in optimal join order, which was not in the reduced power set.
This is a valid and optimal solution, and it does not contain any cartesian product within the solution set.
This example illustrates how ECP reduces the problem complexity by limiting the number of binary variables and constraints and simplifies the QUBO problem to solve with fewer hardware resources.

\subsection{Using ECP-SQSS}
\label{sec:ecp_sqss}

This section discusses how to leverage the advantages of the ECP-SQSS method.
There will be a few changes and new notations as mentioned in section \ref{sec:ecp_theory}.
Although most of the notations remain consistent as in \cite{nayak2024quantum}:
\begin{itemize}
\item $R$ for Relation
\item $C$ for Constraint
\item $\tilde{+}$ for Adding two different subspaces
\item ${{S}^m_{x\tilde{+} y}}$: Where $m = |M|$, and $x$ and $y$ satisfy $x + y = m$.
\item ${P}^m_a$: refers to the subset of joins that include $a$ relations, defined as
\[
P^m_a = \{x \mid x \in P_{ecp} \wedge |x| = a\}
\]
\item Iterative splits are represented as ${{S}^4_{(it(2)\tilde{+} 1)\tilde{+}1}}$, where "it" is used to refer to the iterative nature of the split; therefore, this split sends multiple QUBO problems to solve iteratively.
\item For a split such as ${{S}^7_{(5_t\tilde{+} 1)\tilde{+}1}}$, the subscript $t$ in $5_t$ specifies that it is a total cost variable.
\item $\tilde{T}$ for set of joins with total cost 
\end{itemize}
These notations aim to streamline the integration of the ECP-SQSS, improving the efficiency of QUBO problem-solving.
Further, we will discuss changes in the definitions accordingly.

\subsubsection{Definitions}

\textbf{Definition 1:} The first definition of splitting can be explicitly adapted for ECP, ensuring that it $P^m_a$ utilizes the reduced power set $P_{ecp}$.
Splitting, denoted as ${{S}^m_{x\tilde{+} y}}$ where $x, y \geq 2$, can be defined as:\\
\\
$S^m_{it(a)\tilde{+}it(b)} = \{x\cup y|x\in P^m_a\wedge y=P^m_2\cup...\cup P^m_b -\{z|z\in P^m_2\cup...\cup P^m_b\wedge z\not\subset x\wedge z\cap x\neq \emptyset\}\}$\\
\\
This split takes more than one iteration to solve the whole split, as it is an iterative split.
The use of the ECP method reduces the power set, thereby decreasing the required number of experimental runs.

For instance, consider $m = 5$, $x = 2$, and $y = 3$ where $M = \{R_1, R_2, R_3, R_4, R_5\}$ there is a cross-join between the relations $(R_1, R_2)$ and $(R_1, R_3)$.
In this case, iterations over $(R_1, R_2)$, $(R_1, R_3)$, and $(R_1, R_2, R_3)$ should be discarded, as these intermediate join orders can not be a part of a complete join order.
With ECP, iteration would not occur because these three subspaces have already been eliminated from $P_{ecp}$.

\noindent
\\
\textbf{Definition 2:} The split can be run as an iteration over joins of $a$ relations, where $a < m \text{ and } m = |M|$, such that it covers all possible left/right deep join order joining with the initial intermediate join order of $a$ relations, defined as\\
\\
$S^m_{(it(a)\tilde{+}1\tilde{+}...\tilde{+}1)} = \{x\cup y|y\in P^m_a\wedge x = P^m_{a+1}\cup ...\cup P^m_m - \{z|z\in P^m_{a+1}\cup ...\cup P^m_m\wedge y\not\subset z\}\}$\\
\\
Here $\tilde{+}1$ add those subspaces that subsume the join of $a$ relations on which the split is iterating.
In doing so, it adds all the subspaces till the joins of $m$ relations.
For $a = 2$, it involves single cost binary variables only; for $a > 2$, a split will be used over total cost variables as an initial intermediate join-order.
This definition is still valid when using the ECP method with a reduced power set, $P_{ecp}$.

\noindent
\\
\textbf{Definition 3:} From the first definition of the split ${{S}^m_{x\tilde{+} y}}$, total cost variables would be obtained and can be used for this split to further investigate the bushy join trees. 
This split can be executed iteratively to solve it, and it can be defined as\\
\\
$S^m_{(a_t+b)+c}=\{x\cup y\cup z|x = T^m_a\wedge y=P^m_b\wedge z=P^m_{a+b}\cup P^m_{a+b+c}\}$\\
\\
We will use this definition for our new approach, ECP-SQSS.

\subsection{Reusability of Intermediate Join Order and Punishing Unwanted Subjoins}

For the split $S^8_{it(5)\tilde{+}it(3)}$, the total costs of joining three and five relations for all possible intermediate join orders can be saved.
These total costs are variable; without their corresponding variables (single costs join to form the total cost join order), they can then be utilized while searching for the optimal solution in the splits such as $S^8_{(3_t\tilde{+}4_t)\tilde{+}1}$ and $S^8_{((5_t\tilde{+}1)\tilde{+}1)\tilde{+}1}$, respectively.

In the paper \cite{nayak2024quantum}, it is defined how subjoins, which are not required to come together in a solution set, are split with total cost variables and single cost variables.
Let $\tilde{T}$ be the set of total cost joins for $y\in\tilde{T}$. \footnote{All the other joins $z\in\tilde{P}-\tilde{T}$ contain, as before, the costs of the single join $z$ (without the costs of the subjoins of $z$)}. Subjoins $y$ should be penalized to exclude the costs for these subjoins, and it can be defined as
\begin{equation}
  C_3(\tilde{T},P_{ecp}) = \sum_{y\in\tilde{T}}\hspace{0.2em}\sum_{\forall z\in P_{ecp}:z\subset y} x_y\cdot x_z\cdot w_{max}
\end{equation}

For the split like $S^8_{(5_t\tilde{+}2)\tilde{+}1}$, there is the possibility that the optimal join order chooses more than one variable of two relation joins, but the optimal join order should contain each join variable joining two, five, seven, and eight relations.
Therefore, it needs to penalize the joins of two relations while solving the split, like $S^8_{(5_t\tilde{+}2)\tilde{+}1}$ where the variables joining five relations have total costs. 
Hence, let $x$ and $y$ be the variables joining two relations; hence, we can introduce a fourth constraint such as
\begin{equation}
 C_4(P_{ecp}) = \sum_{\forall x,y\in P_{ecp}:|x|=|y|:x\cap y=\emptyset} x\cdot y \cdot w_{max}
\end{equation}

\subsection{Ensuring Optimal Solutions with the ECP-SQSS}

Splitting the QUBO search space does not result in the loss of the search space where the optimal solution lies.
Instead, splitting offers a trade-off that reduces hardware scalability requirements with the increased number of experimental runs; at the same time, splitting is more promising for finding the optimal solution.
Suppose we divide the search space for a query to join $n$ relations as follows:
\begin{equation}
S = S^n_{(n-1)\tilde{+}1} \cup S^n_{(n-2)\tilde{+}2} \cup \dots \cup S^n_{\lceil n/2 \rceil \tilde{+} \lfloor n/2 \rfloor}
\end{equation}
Here, $S$ represents the entire search space, and the union of all split QUBO subspaces unifies to form the complete QUBO search space.
However, for any $i, j \leq n$, the split search spaces are not necessarily disjoint. That is:
\[
S^n_{i\tilde{+}(n-i)} \cap S^n_{j\tilde{+}(n-j)} \neq \emptyset
\]
To solve problems with higher complexity, leveraging the ECP-SQSS divides the search space into smaller chunks while still maintaining the completeness of the overall search space.
ECP significantly reduces the number of experimental runs required and the hardware resources needed to find the optimal set of solutions.

\section{Experiments}
\label{sec:experiments}
In this section, we will give a description of our local system used to work with the virtual environment for the experiments, and we will also mention the description of the quantum hardware used.
An analysis of the results of the experiments performed for this work.

\subsection{Environment}

Experiments have been performed on gate-based quantum simulators and D-Wave's quantum annealer.
We have used methods such as dynamic programming, SA, QA, VQE, and QAOA. 
For quantum annealing, we have used the Advantage\_system4.1 quantum solver \cite{DWaveQPU_Report}.
It supports 5,627 qubits and 15-way qubit connectivity working at 15.4 ± 1 mK.
D-Wave's Advantage QPU is based on a physical lattice of qubits and couplers known as the Pegasus topology.
As a whole, the Advantage QPU is a lattice of 16x16 such tiles, denoted as a P16 graph.
We have used dwave-neal==0.6.0 to exploit simulated annealing for optimization.
For performing experiments on a gate-based quantum simulator, we have used qiskit == 0.45.1, qiskit-aer == 0.12.0, qiskit-terra == 0.45.1, and qiskit-optimization == 0.5.0.
We have used the local system, an Apple MacBook Pro with an Apple M2 Pro chipset and 16 GB RAM, to perform our experiments.
This device is currently running on macOS 15.2.

For our experiments, we used queries based on the ErgastF1 dataset\footnote{\url{http://ergast.com/mrd/}}, which contains data related to the Formula One series.
The ErgastF1 dataset is sufficiently large to ensure that different join orders exhibit significant variations in processing costs. 
It comprises 13 tables with 19 foreign key relationships between them.

We generated queries involving these tables so that all joins could be performed without requiring a cross-join. 
The queries are constructed using primary key-foreign key relationships as join conditions, with no additional filters applied. 
To retrieve results, we used the \texttt{SELECT *} statement.
Our experiments focus on queries involving 5 to 8 relations, spanning various join graph structures. 

\begin{figure}
\centering
\begin{tikzpicture}
		\node [style=circle,fill=black] (0) at (-4.75, 3.75) {};
		\node [style=circle,fill=black] (1) at (-3.5, 3.5) {};
		\node [style=circle,fill=black] (2) at (-4.25, 2.5) {};
		\node [style=circle,fill=black] (3) at (-2.25, 3.5) {};
		\node [style=circle,fill=black] (4) at (-2.75, 2.5) {};
		\node [style=circle,fill=black] (5) at (-3.25, 4.5) {};
		\node [style=circle,fill=black] (6) at (-0.75, 4) {};
		\node [style=circle,fill=black] (7) at (0, 3.5) {};
		\node [style=circle,fill=black] (8) at (-0.75, 3) {};
		\node [style=circle,fill=black] (9) at (-1.5, 2.5) {};
		\node [style=circle,fill=black] (10) at (1.75, 3.5) {};
		\node [style=circle,fill=black] (11) at (-1.5, 4.5) {};
		\node [style=circle,fill=black] (12) at (1, 3.5) {};
		\node [style=circle,fill=black] (13) at (-4.5, 1.25) {};
		\node [style=circle,fill=black] (14) at (-4.5, 0.5) {};
		\node [style=circle,fill=black] (15) at (-3.25, 1.25) {};
		\node [style=circle,fill=black] (16) at (-3.25, 0.5) {};
		\node [style=circle,fill=black] (17) at (-2.25, 0.5) {};
		\node [style=circle,fill=black] (18) at (-3.25, -0.25) {};
		\node [style=circle,fill=black] (19) at (-0.5, 1.25) {};
		\node [style=circle,fill=black] (20) at (-0.5, 0.5) {};
		\node [style=circle,fill=black] (21) at (0.75, 0.75) {};
		\node [style=circle,fill=black] (22) at (0.75, 0) {};
		\node [style=circle,fill=black] (23) at (2.25, 0.5) {};
		\node [style=circle,fill=black] (24) at (2.25, 1.25) {};
		\node [] (25) at (-3.5, 2.25) {Star};
		\node [] (26) at (0.5, 2.25) {No Cycle};
		\node [] (27) at (-4, -0.75) {Cycle};
		\node [] (28) at (0.75, -0.75) {Multi Cycle};
		\draw (0) to (1);
		\draw (1) to (2);
		\draw (1) to (3);
		\draw (1) to (5);
		\draw (1) to (4);
		\draw (11) to (6);
		\draw (6) to (7);
		\draw (8) to (7);
		\draw (9) to (8);
		\draw (12) to (7);
		\draw (10) to (12);
		\draw (15) to (13);
		\draw (14) to (16);
		\draw (15) to (16);
		\draw (13) to (14);
		\draw (16) to (18);
		\draw (17) to (16);
		\draw (19) to (21);
		\draw (20) to (19);
		\draw (20) to (21);
		\draw (23) to (24);
		\draw (24) to (21);
		\draw (23) to (21);
		\draw (22) to (21);
\end{tikzpicture}

    \caption{Different forms of query graphs}
    \label{fig:graphForms}
\end{figure}
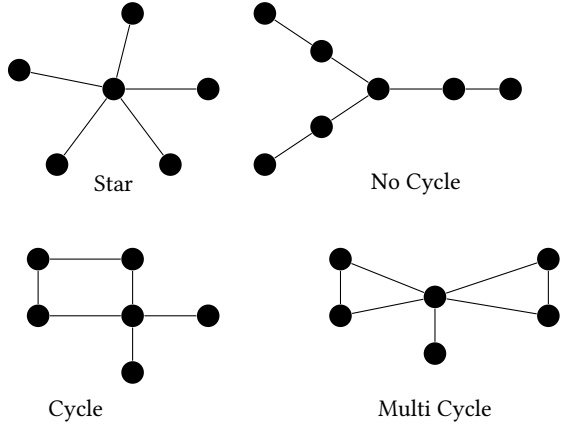

Figure~\ref{fig:graphForms} illustrates the different graph forms, and Table~\ref{tab:queryGraphs} provides a breakdown of the number of queries corresponding to each graph form across different numbers of relations in a query.

\begin{table}[h]
\caption{Graph forms of used queries}
\label{tab:queryGraphs}
\begin{tabular}{|c|c|c|c|c|}
\hline
\textbf{Relations} &  \textbf{Star} &  \textbf{No Cycle}
&  \textbf{Cycle}&  \textbf{\parbox[t]{1cm}{\centering Multi\\ \centering Cycles}}\\
\hline
  \hline
  5   & 21 & 23 & 6 & 0 \\
  \hline
  6   & 15 & 15 & 16 & 3 \\
  \hline
  7   &  0 & 0 & 11 & 14\\
  \hline
  8   & 0 & 0 & 5 & 16 \\
  \hline
  \end{tabular}
\end{table}

We use PostgreSQL (v. 14.4) to calculate the weights required for our QUBO model.
We choose the cardinalities predicted by the internal optimizer of PostgreSQL as our weights, which are also the basis for cost estimations in the PostgreSQL optimizers.
The estimated cardinality of a join is calculated from the estimated cardinalities of the two joined tables and the estimated selectivity of the join conditions \cite{PostgresqlRowEstimation}. 
To calculate these values, PostgreSQL stores the estimated sizes of tables and the number of distinct values, the most common values, and histograms for value distributions for each column.

\subsection{Runtime Complexity of our Approach compared to Exact Methods on Classical Hardware}

Our current implementation uses simple calculations for join cost estimations, but the framework is inherently flexible, allowing for the integration of advanced techniques. 
In principle, we can—and perhaps should—employ high-quality estimations for each join to enhance accuracy.
Traditional cardinality estimation methods offer basic statistical insights but often miss complex data distributions and correlations. 
To overcome these limitations, machine learning-based approaches like the multi-set convolutional network (MSCN) have been developed. 
MSCN captures intricate patterns and accurately predicts join-crossing correlations with which traditional methods struggle \cite{kipf2018learned}.
Incorporating such advanced techniques into our framework can significantly improve cost estimation precision, leading to better join orders. 
This flexibility ensures our approach remains robust across diverse scenarios, supporting both simple and sophisticated estimation methods.

According to Section~\ref{sec:def}, we need a binary variable for each possible subset of the relations with more than 1 relation, excluding the variable for the final join. 
Hence, the size of $\tilde{P}$ is $2^m-m-2$. Recognizing that the structure of the QUBO problem is always the same for the same number to join, except for using other weights, we envision a technique where the QUBO problems can be precompiled to binaries for solving the QUBO problem on a quantum computer. 
Then, for each new join order to optimize, we just replace the $2^m-m-2$ (normalized) weights in the binaries to solve the new QUBO problem on the quantum computer. In this way, the runtime of our approach is $O(2^m-m)$. 

Dynamic programming for join ordering maintains a table $t$ that stores the optimal costs for each possible subjoin and for each relation. The size of this table $t$ is hence $2^m-2$. For each of the entries $S\in P-\{\varnothing\}$ of the dynamic programming table, the dynamic programming approach calculates the optimal costs by $t[S]:=min\{t[S_1]+t[S_2]+cost(S_1\bowtie S_2)|S=S_1 \dot\cup S_2 \wedge S_1\neq \varnothing \wedge S_2\neq \varnothing\}$, where $cost$ is a function to estimate the costs of a given join. Hence, dynamic programming needs $\sum_{i=2}^{m}\binom{m}{i}\cdot(2^{i-1}-1)=\frac{1}{2}\cdot(3^m-2^{m+1}+1)$ comparisons because there are $\binom{m}{i}$ entries for joining $i$ relations, and for each of these entries and for each of these $i$ relations to join, dynamic programming has the option for placing it into $S_1$ or $S_2$ excluding the symmetric cases\footnote{See \cite{10.1145/233269.233317} for the complexity analysis when considering symmetric cases}, $S_1=\varnothing$ and $S_2=\varnothing$.

Our approach avoids the $\sum_{i=2}^{m}\binom{m}{i}\cdot(2^{i-1}-1) = \frac{1}{2}\cdot(3^m-2^{m+1}+1)$ comparisons of dynamic programming, which is the performance advantage when using quantum computing with our approach in comparison to using dynamic programming.

Recently, exact methods on classical hardware have been proposed with worst-case runtime complexities of $O(2^n n^3)$ and $O(2^{3n/2}/\sqrt{\epsilon})$ with an $(1 + \epsilon)$-approximation algorithm for join ordering~\cite{Stoian24DPconv}, but these methods are still not beating our result with a runtime complexity of $O(2^m-m)$.

Furthermore, the number of possible joins is $O(2^m-m)$. Hence, whenever direct costs or cost estimations for each join are used to determine the best join order, then our runtime complexity $O(2^m-m)$ is theoretically optimal, i.e., our worst-case runtime complexity meets the theoretical lowest possible runtime complexity for exact methods solving the join ordering problem.

\subsection{Number Of Required Qubits}
\label{sec:qubit_requirement}
In our previous work \cite{nayak2024quantum}, we made an attempt to solve a complex problem as a subproblem in order to overcome hardware limitations with limited scalability.
Although there are use cases, such as when there are cross-joins present in the dataset, where direct use of this method may not be a wise choice.
There is a possibility that finding the optimal solution may not require as many qubits as it required in our previous work.
In our proposed ECP-SQSS method in section \ref{sec:ecp_theory}, the required qubits have been further significantly reduced with the reduced binary variables in the power set $(P_{ecp})$.
The reduced power set $P_{ecp}$ is defined by eliminating the tables that do not share any foreign key relationship and that the incoming query requires to process.
Consequently, a study on the required number of qubits for varying numbers of cross-join pairs between the dataset relations is both reasonable and insightful.

For instance, consider that we are solving the join-ordering problem for a query containing four relations $\{R_1, R_2, R_3, R_4\}$. 
If $R_1$, $R_2$, and $R_3$  share foreign key relationships among themselves, any combination of these three relations will not result in a cartesian product.
However, if it $R_4$  shares a foreign key relationship only with $R_3$ but not with $R_1$ or $R_2$, there will be two possible combinations of cross-joins between the relations when searching for the optimal execution plan, such as $(R_1, R_4)$ and $(R_2, R_4)$.
These two pairs of relations will influence the intermediate results, such that the intermediate result joining $\{R_1, R_2, R_4\}$ will be invalid, as $R_4$  does not have a direct foreign key relationship with $R_1$ and $R_2$. Therefore, in order to reduce the joining cost of the relations, we can discard the intermediate result $\{R_1, R_2, R_4\}$ from the power set.
Now, the resulting join ordering problem would require fewer qubits to solve.

\begin{figure}[htb!]
\centering
\begin{tikzpicture}
\begin{axis}[
legend style={nodes={scale=0.6, transform shape}},
xmin = 0, xmax = 14,
ymin = 0, ymax = 100,
xtick distance = 1,
ytick distance = 10,
grid = major,
width = \columnwidth,
height = 0.4\textwidth,
xlabel = {Number of cross-join pairs},
ylabel = {Reduction in required qubits(\%)},
]
\addplot [color=orange, mark=diamond*] table [x = num_of_CJ, y = reducn_qubits_perc, col sep=comma,]{req_qubits_R3.csv};
\addlegendentry{Joining 3 relations}

\addplot [color=magenta, mark=halfsquare*] table [x = num_of_CJ, y = reducn_qubits_perc, col sep=comma,]{req_qubits_R4.csv};
\addlegendentry{Joining 4 relations}

\addplot [color=red, mark=triangle*] table [x = num_of_CJ, y = reducn_qubits_perc, col sep=comma,]{req_qubits_R5.csv};
\addlegendentry{Joining 5 relations}

\addplot [color=teal, mark=star] table [x = num_of_CJ, y = reducn_qubits_perc, col sep=comma,]{req_qubits_R6.csv};
\addlegendentry{Joining 6 relations}

\addplot [color=olive, mark=halfdiamond*] table [x = num_of_CJ, y = reducn_qubits_perc, col sep=comma,]{req_qubits_R7.csv};
\addlegendentry{Joining 7 relations}

\addplot [color= magenta, mark= -, thick, mark size=4pt] table [x = num_of_CJ, y = reducn_qubits_perc, col sep=comma,]{req_qubits_R4_cj3.csv};
\addplot [color= magenta, mark= halfsquare*, thick] table [x = num_of_CJ, y = reducn_qubits_perc, col sep=comma,]{req_qubits_R4_cj3.csv};

\addplot [color= red, mark= -, thick, mark size=4pt] table [x = num_of_CJ, y = reducn_qubits_perc, col sep=comma,]{req_qubits_R5_cj5.csv};
\addplot [color= red, mark= triangle*, thick] table [x = num_of_CJ, y = reducn_qubits_perc, col sep=comma,]{req_qubits_R5_cj5.csv};
\addplot [color= red, mark= -, thick, mark size=4pt] table [x = num_of_CJ, y = reducn_qubits_perc, col sep=comma,]{req_qubits_R5_cj6.csv};
\addplot [color= red, mark= triangle*, thick] table [x = num_of_CJ, y = reducn_qubits_perc, col sep=comma,]{req_qubits_R5_cj6.csv};

\addplot [color= teal, mark= -, thick, mark size=4pt] table [x = num_of_CJ, y = reducn_qubits_perc, col sep=comma,]{req_qubits_R6_cj4.csv};
\addplot [color= teal, mark= star] table [x = num_of_CJ, y = reducn_qubits_perc, col sep=comma,]{req_qubits_R6_cj4.csv};
\addplot [color= teal, mark= -, thick, mark size=4pt] table [x = num_of_CJ, y = reducn_qubits_perc, col sep=comma,]{req_qubits_R6_cj5_minmax.csv};
\addplot [color= teal, mark= star] table [x = num_of_CJ, y = reducn_qubits_perc, col sep=comma,]{req_qubits_R6_cj5.csv};
\addplot [color= teal, mark= -, thick, mark size=4pt] table [x = num_of_CJ, y = reducn_qubits_perc, col sep=comma,]{req_qubits_R6_cj6.csv};
\addplot [color= teal, mark= star] table [x = num_of_CJ, y = reducn_qubits_perc, col sep=comma,]{req_qubits_R6_cj6.csv};
\addplot [color= teal, mark= -, thick, mark size=4pt] table [x = num_of_CJ, y = reducn_qubits_perc, col sep=comma,]{req_qubits_R6_cj7.csv};
\addplot [color= teal, mark= star] table [x = num_of_CJ, y = reducn_qubits_perc, col sep=comma,]{req_qubits_R6_cj7.csv};
\addplot [color= teal, mark= -, thick, mark size=4pt] table [x = num_of_CJ, y = reducn_qubits_perc, col sep=comma,]{req_qubits_R6_cj8_minmax.csv};
\addplot [color= teal, mark= star] table [x = num_of_CJ, y = reducn_qubits_perc, col sep=comma,]{req_qubits_R6_cj8.csv};
\addplot [color= teal, mark= -, thick, mark size=4pt] table [x = num_of_CJ, y = reducn_qubits_perc, col sep=comma,]{req_qubits_R6_cj9.csv};
\addplot [color= teal, mark= star] table [x = num_of_CJ, y = reducn_qubits_perc, col sep=comma,]{req_qubits_R6_cj9.csv};

\addplot [color= olive, mark= -, thick, mark size=4pt] table [x = num_of_CJ, y = reducn_qubits_perc, col sep=comma,]{req_qubits_R7_cj5.csv};
\addplot [color= olive, mark= halfdiamond*, thick] table [x = num_of_CJ, y = reducn_qubits_perc, col sep=comma,]{req_qubits_R7_cj5.csv};
\addplot [color= olive, mark= -, thick, mark size=4pt] table [x = num_of_CJ, y = reducn_qubits_perc, col sep=comma,]{req_qubits_R7_cj6.csv};
\addplot [color= olive, mark= halfdiamond*, thick] table [x = num_of_CJ, y = reducn_qubits_perc, col sep=comma,]{req_qubits_R7_cj6.csv};
\addplot [color= olive, mark= -, thick, mark size=4pt] table [x = num_of_CJ, y = reducn_qubits_perc, col sep=comma,]{req_qubits_R7_cj7.csv};
\addplot [color= olive, mark= halfdiamond*, thick] table [x = num_of_CJ, y = reducn_qubits_perc, col sep=comma,]{req_qubits_R7_cj7.csv};
\addplot [color= olive, mark= -, thick, mark size=4pt] table [x = num_of_CJ, y = reducn_qubits_perc, col sep=comma,]{req_qubits_R7_cj9_minmax.csv};
\addplot [color= olive, mark= halfdiamond*, thick] table [x = num_of_CJ, y = reducn_qubits_perc, col sep=comma,]{req_qubits_R7_cj9.csv};
\addplot [color= olive, mark= -, thick, mark size=4pt] table [x = num_of_CJ, y = reducn_qubits_perc, col sep=comma,]{req_qubits_R7_cj10_minmax.csv};
\addplot [color= olive, mark= halfdiamond*, thick] table [x = num_of_CJ, y = reducn_qubits_perc, col sep=comma,]{req_qubits_R7_cj10.csv};
\addplot [color= olive, mark= -, thick, mark size=4pt] table [x = num_of_CJ, y = reducn_qubits_perc, col sep=comma,]{req_qubits_R7_cj11.csv};
\addplot [color= olive, mark= halfdiamond*, thick] table [x = num_of_CJ, y = reducn_qubits_perc, col sep=comma,]{req_qubits_R7_cj11.csv};

\end{axis}
\end{tikzpicture}
\caption{This figure shows the variation of qubit requirements on the varying complexity of the query. For a query to join any number of relations, for a particular value of cross-join pairs depicted on the x-axis, which can have multiple values for its corresponding y-axis (the reduction(\%) in qubit requirement), which has also been shown in the figure by vertical marks.}
\label{fig:ecp_qubits_required}
\end{figure}
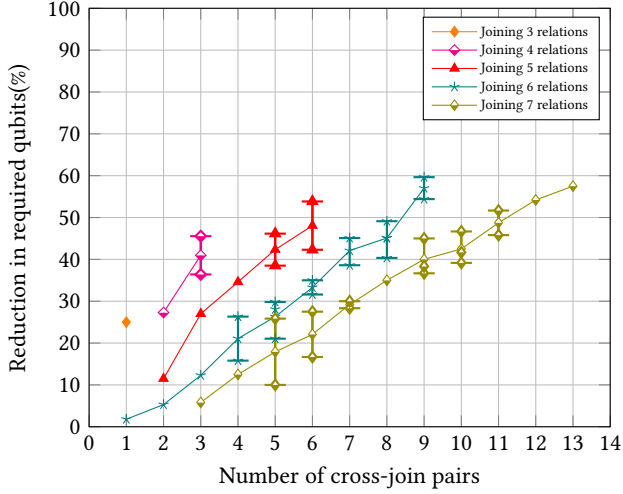

In our previous work, we did an analysis for the required number of qubits by varying the problem size, as shown in Table~\ref{tab:NrOfQubits}.
Similarly, we did the same analysis for the ECP method to find out the reduction in qubit requirements to solve the same QUBO problem in figure~\ref{fig:ecp_qubits_required}.
\begin{table}[h]
\caption{Number of required qubits}
\label{tab:NrOfQubits}
\begin{tabular}{|c|c|}
\hline
\textbf{Number of relations} &  \textbf{Number of qubits}\\
\hline
  3   & 4  \\
  \hline
  4    & 11 \\
  \hline
  5   & 26 \\
  \hline
  6   & 57 \\
  \hline
  7   & 120 \\
  \hline
  \end{tabular}
\end{table}

Although the required qubits do not explicitly depend on the count of cross-joins between the relations present in the query.
It can vary even with the same data shown in figure~\ref{fig:ecp_qubits_required}.
So it also depends on relations that appeared more than once in the cross joins.
On the continuing above example, we have two cross-join pairs as: $\{(R_1, R_4),(R_2, R_4)\}$, therefore, the eliminated intermediate results will be $\{\{R_1, R_4\},\{R_2, R_4\},\{R_1, R_2, R_4\}\}$ but if a query contains two pairs of cross-joins as: $\{(R_1, R_2),(R_3, R_4)\}$ then eliminated intermediate results will be $\{\{R_1, R_2\},\{R_3, R_4\}\}$ only, but the number of cross-joins is same for both cases.

From the above, we can define the minimum and maximum number of intermediate cartesian joins that can be eliminated.
For a query with $m$ relations, the minimum number of eliminated intermediate cartesian joins is equal to the number of cross-join pairs if each relation in the query has a cross-join with exactly one relation.
However, the maximum number of eliminations of binary variables occurs when each relation in the query has a cross-join with exactly $m-2$ relations.

\subsection{Required experimental runs}
\label{subsec:required-exp-runs}
\begin{figure}[htb!]
\centering
\begin{tikzpicture}
\begin{axis}[
legend style={nodes={scale=0.6, transform shape}},
legend pos=south east,
xmin = 0, xmax = 14,
ymin = 0, ymax = 100,
xtick distance = 1,
ytick distance = 10,
grid = major,
width = \columnwidth,
height = 0.4\textwidth,
xlabel = {Number of cross-join pairs},
ylabel = {Reduced experimental runs(\%)},
]

\addplot [color=orange, mark=square] table [x = num_of_CJ, y = reducn_exp_runs_perc, col sep=comma,]{req_qubits_R4.csv};
\addlegendentry{Joining 4 relations}

\addplot [color=red, mark=star] table [x = num_of_CJ, y = reducn_exp_runs_perc, col sep=comma,]{req_qubits_R5.csv};
\addlegendentry{Joining 5 relations}

\addplot  [color=teal, mark=o] table [x = num_of_CJ, y = reducn_exp_runs_perc, col sep=comma,]{req_qubits_R6.csv};
\addlegendentry{Joining 6 relations}

\addplot  [color=olive, mark=halfsquare*] table [x = num_of_CJ, y = reducn_exp_runs_perc, col sep=comma,]{req_qubits_R7.csv};
\addlegendentry{Joining 7 relations}

\addplot [color= red, mark= -, thick, mark size=4pt] table [x = num_of_CJ, y = reducn_exp_runs_perc, col sep=comma,]{req_qubits_R5_cj5.csv};
\addplot [color= red, mark= star] table [x = num_of_CJ, y = reducn_exp_runs_perc, col sep=comma,]{req_qubits_R5_cj5.csv};
\addplot [color= red, mark= -, thick, mark size=4pt] table [x = num_of_CJ, y = reducn_exp_runs_perc, col sep=comma,]{req_qubits_R5_cj6.csv};
\addplot [color= red, mark= star] table [x = num_of_CJ, y = reducn_exp_runs_perc, col sep=comma,]{req_qubits_R5_cj6.csv};

\addplot [color= teal, mark= -, thick, mark size=4pt] table [x = num_of_CJ, y = reducn_exp_runs_perc, col sep=comma,]{req_qubits_R6_cj4.csv};
\addplot [color= teal, mark= o] table [x = num_of_CJ, y = reducn_exp_runs_perc, col sep=comma,]{req_qubits_R6_cj4.csv};
\addplot [color= teal, mark= -, thick, mark size=4pt] table [x = num_of_CJ, y = reducn_exp_runs_perc, col sep=comma,]{req_qubits_R6_cj5_minmax_exp.csv};
\addplot [color= teal, mark= o] table [x = num_of_CJ, y = reducn_exp_runs_perc, col sep=comma,]{req_qubits_R6_cj5.csv};
\addplot [color= teal, mark= -, thick, mark size=4pt] table [x = num_of_CJ, y = reducn_exp_runs_perc, col sep=comma,]{req_qubits_R6_cj9.csv};
\addplot [color= teal, mark= o] table [x = num_of_CJ, y = reducn_exp_runs_perc, col sep=comma,]{req_qubits_R6_cj9.csv};

\addplot [color= olive, mark= -, thick, mark size=4pt] table [x = num_of_CJ, y = reducn_exp_runs_perc, col sep=comma,]{req_qubits_R7_cj6.csv};
\addplot [color= olive, mark= halfsquare*] table [x = num_of_CJ, y = reducn_exp_runs_perc, col sep=comma,]{req_qubits_R7_cj6.csv};
\addplot [color= olive, mark= halfsquare*] table [x = num_of_CJ, y = reducn_exp_runs_perc, col sep=comma,]{req_qubits_R7_cj7.csv};
\addplot [color= olive, mark= -, thick, mark size=4pt] table [x = num_of_CJ, y = reducn_exp_runs_perc, col sep=comma,]{req_qubits_R7_cj9_minmax_exp.csv};
\addplot [color= olive, mark= halfsquare*] table [x = num_of_CJ, y = reducn_exp_runs_perc, col sep=comma,]{req_qubits_R7_cj9.csv};
\addplot [color= olive, mark= -, thick, mark size=4pt] table [x = num_of_CJ, y = reducn_exp_runs_perc, col sep=comma,]{req_qubits_R7_cj10_minmax.csv};
\addplot [color= olive, mark= halfsquare*] table [x = num_of_CJ, y = reducn_exp_runs_perc, col sep=comma,]{req_qubits_R7_cj10.csv};
\addplot [color= olive, mark= -, thick, mark size=4pt] table [x = num_of_CJ, y = reducn_exp_runs_perc, col sep=comma,]{req_qubits_R7_cj11.csv};
\addplot [color= olive, mark= halfsquare*] table [x = num_of_CJ, y = reducn_exp_runs_perc, col sep=comma,]{req_qubits_R7_cj11.csv};

\end{axis}
\end{tikzpicture}
\caption{Number of runs for experiments to be done for our new method for queries of different numbers of relations. We compared it with the SQSS, which required a fixed number of experiments for each query, while the ECP-SQSS requires runs of experiments according to the number of cross-joins for a given query. It can have multiple values along the y-axis for a given number of cross-joins in the given query, which has been depicted by vertical marks.}
\label{fig:exp_runs_ecp_sqss}
\end{figure}
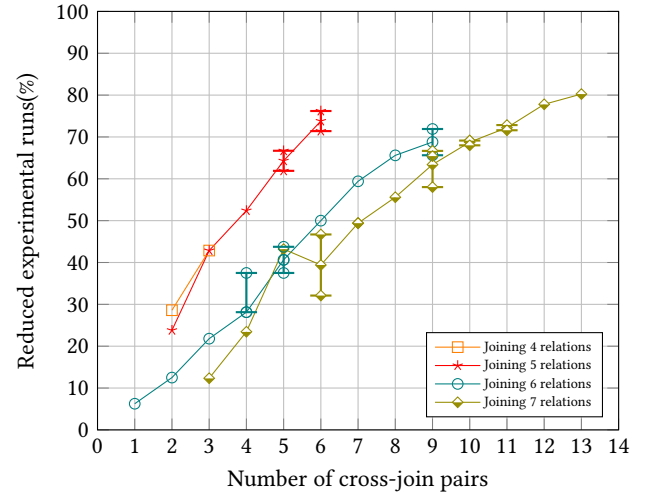

Experimental runs required for a query of the respective number of relations are mentioned in Table~\ref{tab:NumberOfExperiments}. 
The required number of total experimental runs depends on whether the split is iterative or we have chosen it to be completed in a single run, if it is feasible to run on the quantum device.
Hence, we have split the QUBO search space for queries of different sizes in order to observe the variation in the qubits' requirements and required experimental runs.
In fig~\ref{fig:exp_runs_ecp_sqss}, we have compared the experimental runs required by the proposed ECP-SQSS method in this work with the total experimental runs to solve the QUBO problem in the previous work~\cite{nayak2024quantum}.

\begin{table}[htb!]
\caption{Number of experimental runs required to successfully solve the join ordering problem on D-Wave's quantum annealer for the varying number of relations for a given query with the previous method SQSS.}
\label{tab:NumberOfExperiments}
\begin{tabular}{|c|c|}
\hline
\textbf{Number of relations} &  \textbf{Number of Exp. run}\\
  \hline
  4  & 7 \\
  \hline
  5  & 21 \\
  \hline
  6  & 32  \\
  \hline
  7  & 81  \\
\hline
\end{tabular}
\end{table}

Required experimental runs for our experiments while employing SQSS and eliminating the Cartesian product, showed variation with varying the number of cross-join pairs present in the query. 
Benchmarking the effect of the method with the real-world dataset ErgastF1 has been illustrated in Figure~\ref{fig:exp_runs_ecp_sqss}.
The required experimental runs may also vary for the same number of cross-joins as depicted in Figure~\ref{fig:exp_runs_ecp_sqss} with a different number of required qubits.
Such variation in the required qubits arises for different SQL queries in the dataset with a similar number of cross-joins present.
In figure~\ref{fig:ecp_qubits_required} and figure~\ref{fig:exp_runs_ecp_sqss}, it is also clear that for two different queries running on a particular dataset having the same number of cross-joins may require a different number of qubits for running the experiment; the required number of experimental runs may also vary for the same dataset.

\subsection{Result Analysis}
\label{subsec:result-analysis}
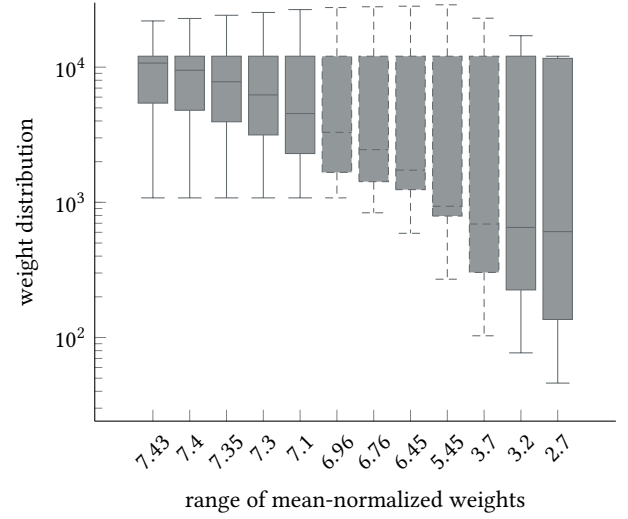
\begin{figure}[htb!]
\centering
    \begin{tikzpicture}
		\begin{axis}[width=\linewidth,
		boxplot/draw direction=y,
		axis x line*=bottom,
		axis y line*=left,
		ymin=0,
		ymax=30000,
		xtick={1,2,3,4,5,6,7,8,9,10,11,12},
            xticklabels={7.43, 7.4, 7.35, 7.3, 7.1, 6.96, 6.76, 6.45, 5.45, 3.7, 3.2, 2.7},
            xticklabel style={rotate=45},
		ylabel={weight distribution},
            xlabel={range of mean-normalized weights},
            width = \columnwidth,
            height = 0.4\textwidth,
            ymode=log
		]
            \addplot+[color=Diagramm1,fill=Diagramm1!70,boxplot prepared={
			lower whisker=1079, lower quartile=5423,
			median=10721, upper quartile=12050,
			upper whisker=21990.5}]
		coordinates {};
            \addplot+[color=Diagramm1,fill=Diagramm1!70,boxplot prepared={
			lower whisker=1079, lower quartile=4801.25,
			median=9491, upper quartile=12050,
			upper whisker=22923}]
		coordinates {};
            \addplot+[color=Diagramm1,fill=Diagramm1!70,boxplot prepared={
			lower whisker=1079, lower quartile=3946,
			median=7801, upper quartile=12050,
			upper whisker=24206}]
		coordinates {};
            \addplot+[color=Diagramm1,fill=Diagramm1!70,boxplot prepared={
			lower whisker=1079, lower quartile=3154,
			median=6235, upper quartile=12050,
			upper whisker=25394}]
		coordinates {};
            \addplot+[color=Diagramm1,fill=Diagramm1!70,boxplot prepared={
			lower whisker=1079, lower quartile=2296.75,
			median=4540, upper quartile=12050,
			upper whisker=26680}]
		coordinates {};
            \addplot+[color=Diagramm1,fill=Diagramm1!70,boxplot prepared={
			lower whisker=1079, lower quartile=1666.5,
			median=3294, upper quartile=12050,
			upper whisker=27625.25}]
		coordinates {};
            \addplot+[color=Diagramm1,fill=Diagramm1!70,boxplot prepared={
			lower whisker=837, lower quartile=1422.75,
			median=2454, upper quartile=12050,
			upper whisker=27991}]
		coordinates {};
            \addplot+[color=Diagramm1,fill=Diagramm1!70,boxplot prepared={
			lower whisker=590, lower quartile=1242,
			median=1731, upper quartile=12050,
			upper whisker=28262}]
		coordinates {};
            \addplot+[color=Diagramm1,fill=Diagramm1!70,boxplot prepared={
			lower whisker=270, lower quartile=792,
			median=935.5, upper quartile=12050,
			upper whisker=28937}]
		coordinates {};
            \addplot+[color=Diagramm1,fill=Diagramm1!70,boxplot prepared={
			lower whisker=103, lower quartile=303,
			median=691, upper quartile=12050,
			upper whisker= 23028}]
		coordinates {};
            \addplot+[color=Diagramm1,fill=Diagramm1!70,boxplot prepared={
			lower whisker=77, lower quartile=225,
			median=652, upper quartile=12050,
			upper whisker= 17100}]
		coordinates {};
            \addplot+[color=Diagramm1,fill=Diagramm1!70,boxplot prepared={
			lower whisker=46, lower quartile=136,
			median=607.5, upper quartile=11621.5,
			upper whisker=12050}]
		coordinates {};
		\end{axis}
		\end{tikzpicture}
		\caption{This boxplot illustrates weight distribution assigned to the binary variables of QUBO with the variation in the range of the mean normalized of the weights. As the range decreases, the number of retrieved optimal shots increases using the SQSS approach, corresponding to a decrease in selectivity. For example, in a query involving five relations, the predicate constructorstandings.raceId $> k$ is used, where ‘constructorstandings’ represents a relation in the query, ‘raceId’ is a column in the relation, and $k$, which should satisfy the predicate in order to decrease the selectivity.}
		\label{fig:BoxPlot_WD_r5_q0}
\end{figure}

In this section, we will discuss the experimental results and their related statistics, like optimal shot percentage, consistency of obtaining the solution, weight distribution, and range of mean-normalized weight.
Before we dive in we need to understand what these statistical terms exactly represent. 
Optimal shot percentage can be defined as the percentage of obtained optimal shots that represents our optimal solution of the query when solving it using quantum annealing on D-Wave's quantum annealer.
Consistency of obtaining the solution can basically be defined as how consistently the optimal solution of the queries with different complexities is obtained when solving them.
Weight distribution is simply the variation in the weight, which has been plotted using a box plot in figure~\ref{fig:BoxPlot_WD_r5_q0}.
The range of the mean-normalized weight is defined for the set of weights assigned to the binary variable of the join ordering QUBO problem at hand to solve.
First, we normalized each weight with the mean of the weights and then took the difference of the maximum and minimum values from the normalized value as used in figure~\ref{fig:BoxPlot_WD_r5_q0}.

Mathematically, for a given weight of the binary variables in an iteration of the QUBO split the range of mean normalized values has been calculated by following the steps given below.

Let the input be:
\[
E_{itr} = \{w_1, w_2, \dots, w_n\}, \quad w_j \in \mathbb{R}, \quad 1 \leq j \leq n
\]
Where $w_j$ is the weight of binary variables for each experimental iteration ($E_{itr}$) of the split having $n$ binary variables.
For each iteration:\\
1. Compute the mean: 
   \[
   \mu = \frac{1}{n} \sum_{j=1}^{n} w_j
   \]
2. Compute the normalized weights:
   \[
   \tilde{E_{itr}} = \frac{w_j}{\mu}, \quad \forall w_j \in E_{itr}
   \]
3. Compute the range of mean normalized values:
   \[
   \text{range} = \max(\tilde{E_{itr}}) - \min(\tilde{E_{itr}})
   \]

\begin{figure}[htb!]
\centering
\begin{tikzpicture}
\begin{axis}[legend style={nodes={scale=0.6, transform shape}},
xmin = 5, xmax = 8,
ymin = 0, ymax = 0.8,
xtick distance = 1,
ytick distance = 0.1,
grid = both,
minor tick num = 1,
width = \columnwidth,
height = 0.4\textwidth,
xlabel = {Number of relations},
ylabel = {Fraction of optimal shots},
]
\addplot [color=blue, mark=oplus] table [x = relations, y = frac_optimal_shots, col sep=comma,]{SA_fracn_optimal_shots.csv};
\addlegendentry{SA with ECP-SQSS}
\addplot [color=cyan, mark=halfsquare*] table [x = relations, y = frac_optimal_shots, col sep=comma,]{QA_fracn_optimal_shots.csv};
\addlegendentry{QA with ECP-SQSS}
\addplot [color=red, mark=square] table [x = relations, y = prev_optimal_shots, col sep=comma,]{SA_fracn_optimal_shots.csv};
\addlegendentry{SA with SQSS}
\addplot [color=black, mark=o] table [x = relations, y = prev_optimal_shots, col sep=comma,]{QA_fracn_optimal_shots.csv};
\addlegendentry{QA with SQSS}
\end{axis}
\end{tikzpicture}
\caption{The fraction of the optimal shots' variation with the number of relations of queries to get the optimal join order using simulated annealing and quantum annealing.}
\label{optimal_shots_using_annealing}
\end{figure}
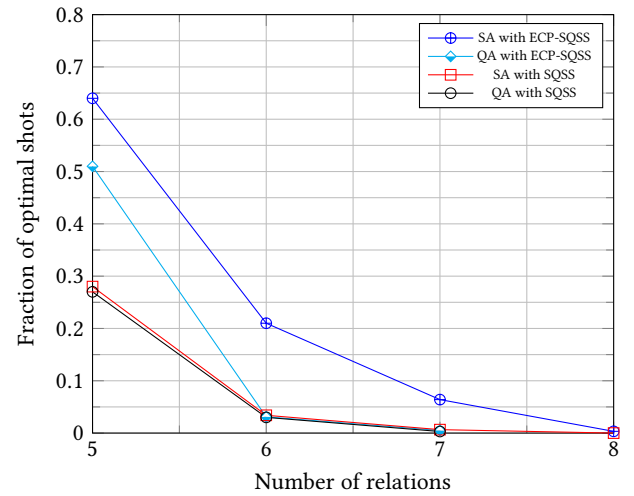

Our study has been benchmarked using real-world SQL queries on the ErgastF1 dataset for the QA and SA methods.
On running experiments on D-Wave's Quantum Annealer, optimal shots obtained in the solution set for the queries that include three and four relations are $100\%$ and $99.95\%$, respectively.
The comparison of obtained optimal shots, depicted in figure~\ref{optimal_shots_using_annealing}, highlights the improved performance of the study on the ECP-SQSS approach over the previous SQSS method. 
This improvement underscores the effectiveness of incorporating ECP, particularly in scenarios in which join-order performs a cross-join between the relations.
In the case of SQL queries that include five relations' queries, the optimal shots retrieved on solving using quantum annealing have nearly doubled, achieving $51.14\%$ 1000 optimal shots out of 1000 total shots on D-Wave's quantum annealer, compared to $27.06\%$ with the previous method for the tested dataset ErgastF1.
For queries involving six and seven relations, the current approach achieved $3.13\%$ and $0.5\%$ optimal results, respectively, highlighting its limitations as the complexity of the queries increases.

On experimenting with SA for our new study, we observed that it achieved $99.01\%$ and $93.94\%$ of optimal shots for three and four relations, respectively, out of a total of $100000$ shots.
This suggests that quantum annealers tend to perform exceptionally well for small to medium-sized problems when we compare the results obtained for queries of three and four relations using QA and SA.
On the other hand, we obtained $64.09\%$ of the optimal shots for queries including five relations, whereas previously, the SQSS method retrieved $27.868\%$ optimal shots.
For a query joining six relations, it obtained $21.2\%$ optimal shots, and $6.4\%$ for a query joining seven relations, it did.
We also obtained $0.33\%$ optimal shots using SA for queries of eight relations, but QA failed with the higher complexity of the problems.
This outcome highlights the limitations of the current state-of-the-art using the quantum annealing method.
Incorporating ECP with SQSS significantly enhanced the performance of QA and SA for queries, especially those that consist of five relations, due to a further reduction of the number of variables and constraints in the QUBO problem, enabling us to solve the QUBO problems of real-world queries of higher complexity.

\begin{figure}[htb!]
     \centering
    \begin{tikzpicture}[scale=0.8, transform shape]
        \begin{axis}[%
        every axis plot post/.style={/pgf/number format/fixed},
             ylabel shift = -1em,
             ybar,
             ybar=1pt,
             bar width=10pt,
             x=2.2cm,
             xmin=2.5,
             xmax = 5.5,
             ymin=50,
             ymax=100,
            axis on top,
             xtick=data,
             legend columns=6,
             legend style={cells={align=left},at={(0.45,-0.2)},anchor=north},
             xlabel={Number of relations},
             ylabel style={align=center},
             ylabel={Consistency(\%)},
             tick align = outside,
             axis lines*=left,
             ]
        \addlegendimage{empty legend}
        \addlegendentry{\textbf{QAOA:}}

        \addplot table[x=relation,y=perctg, col sep=comma,text special chars={\#}]{QAOA_cobyla.csv};
        \addlegendentry{COBYLA}
         
        \addplot table[x=relation,y=perctg, col sep=comma,text special chars={\#}]{QAOA_spsa.csv};
        \addlegendentry{SPSA}
         
        \addlegendimage{empty legend}
        \addlegendentry{\textbf{VQE:}}

        \addplot table[x=relation,y=perctg, col sep=comma,text special chars={\#}]{SVQE_cobyla.csv};
        \addlegendentry{COBYLA}

        \addplot table[x=relation,y=perctg, col sep=comma,text special chars={\#}]{SVQE_spsa.csv};
        \addlegendentry{SPSA}

        \end{axis}
    \end{tikzpicture}
\caption{The figure shows the percentage of real-world queries for a given number of queries for which optimal join order has been retrieved on the universal quantum simulator.}
\label{fig:universal_quantum_simulator_ecp_sqss}
\end{figure}
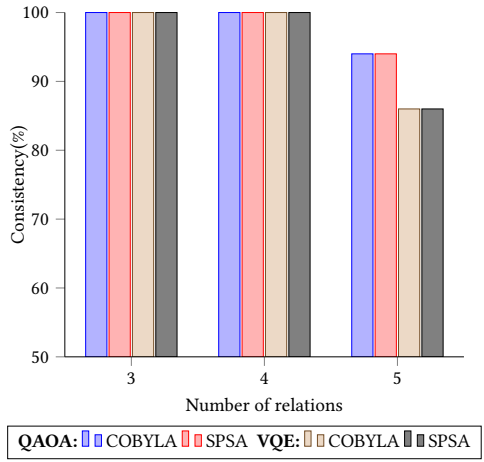

We tested real-world queries on the universal quantum simulator to study eliminating cross-joins from intermediate query plans with SQSS. 
Test results are shown in figure~\ref{fig:universal_quantum_simulator_ecp_sqss}.
Experiments on the simulator for queries including three and four relations obtained optimal join order with $100\%$ consistency for the set of queries in our study on employing the variational quantum algorithms like QAOA and VQE.
The performance of these variational algorithms on the simulator is nearly close to that $100\%$ of finding the join order of queries, including five relations using the QAOA algorithm; on the other hand, it shows less consistency on solving the problems with the VQE algorithm, which is $86\%$.
These results show significant improvement on the quantum simulator when excluding cross-joins.

The following is a SQL query that joins five relations, used for the weight distribution analysis in the figure~\ref{fig:BoxPlot_WD_r5_q0}.
\begin{lstlisting}[language = SQL,
           showspaces=false,
           basicstyle=\ttfamily,
           commentstyle=\color{gray}]
SELECT COUNT(*) FROM 
circuits,
constructorresults,
constructors,
constructorstandings,
races
WHERE 
constructorresults.constructorId=constructors.constructorId
AND constructorresults.raceId=races.raceId
AND constructorstandings.constructorId=constructors.constructorId
AND constructorstandings.raceId=races.raceId
AND races.circuitId=circuits.circuitId
AND constructorstandings.raceId > k;
\end{lstlisting}

\begin{figure}[htb!]
\centering
\begin{tikzpicture}
\begin{axis}[legend style={nodes={scale=0.65, transform shape}},
xmin = 1, xmax = 8,
ymin = 10000, ymax = 100000,
xtick distance = 1,
ytick distance = 10000,
grid = both,
xlabel style={font=\small},        
ylabel style={font=\small, yshift=-3pt},  
tick label style= {font=\small},
minor tick num = 1,
width = \columnwidth,
height = 0.4\textwidth,
xlabel = {Range of mean-normalized weights},
ylabel = {Number of Optimal shots},
]
\addplot [color=red, mark=square] table [x = diff_weight, y = optimal_shots, col sep=comma,]{diff_weight_sss_sa_r5_q0.csv};
\addlegendentry{query-1}
\addplot [color=orange, mark=o] table [x = diff_weight, y = optimal_shots, col sep=comma,]{diff_weight_sss_sa_r5_q4.csv};
\addlegendentry{query-2}
\addplot [color=olive, mark=halfsquare*] table [x = diff_weight, y = optimal_shots, col sep=comma,]{diff_weight_sss_sa_r5_q6.csv};
\addlegendentry{query-3}
\addplot [color=blue, mark=diamond*] table [x = diff_weight, y = optimal_shots, col sep=comma,]{diff_weight_sss_sa_r5_q7.csv};
\addlegendentry{query-4}
\addplot [color=black, mark=diamond] table [x = diff_weight, y = optimal_shots, col sep=comma,]{diff_weight_sss_sa_r5_q8.csv};
\addlegendentry{query-5}
\addplot [color=teal, mark=star] table [x = diff_weight, y = optimal_shots, col sep=comma,]{diff_weight_sss_sa_r5_q53.csv};
\addlegendentry{query-6}
\addplot [color=red, mark=+] table [x = diff_weight, y = optimal_shots, col sep=comma,]{diff_weight_sss_sa_r5_q159.csv};
\addlegendentry{query-7}
\addplot [color=olive, mark=asterisk] table [x = diff_weight, y = optimal_shots, col sep=comma,]{diff_weight_sss_sa_r5_q202.csv};
\addlegendentry{query-8}
\addplot [color=teal, mark=triangle] table [x = diff_weight, y = optimal_shots, col sep=comma,]{diff_weight_sss_sa_r5_q271.csv};
\addlegendentry{query-9}
\addplot [color=black, mark=triangle*] table [x = diff_weight, y = optimal_shots, col sep=comma,]{diff_weight_sss_sa_r5_q5.csv};
\addlegendentry{query-10}
\end{axis}
\end{tikzpicture}
\caption{The above figure shows the variation of the number of optimal shots with the range of mean-normalized weights for the experiments using simulated annealing. All these experiments have been performed above for the split that consists of eight binary variables in the QUBO search space for the query of five relations.}
\label{fig:optimal_shots_sa_mean-normalised_weight_range}
\end{figure}
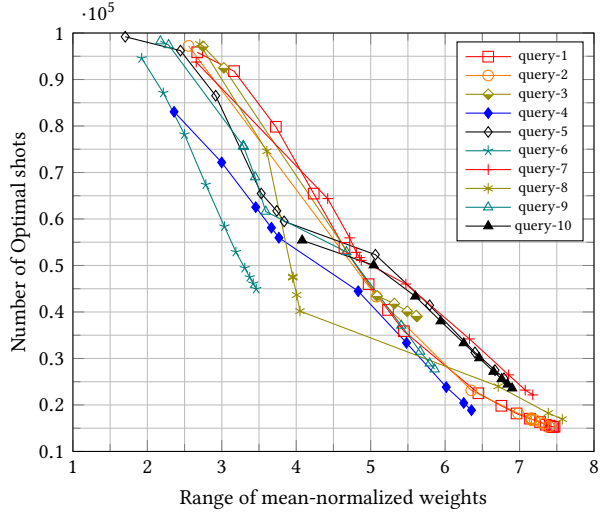

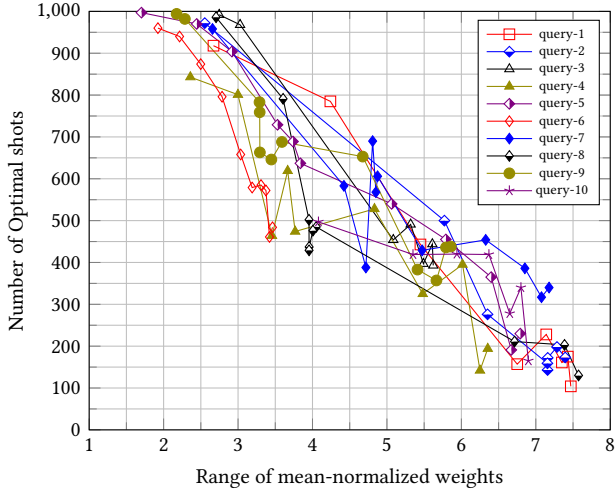
\begin{figure}[htb!]
\centering
\begin{tikzpicture}
\begin{axis}[legend style={nodes={scale=0.6, transform shape}}, 
xmin = 1, xmax = 8,
ymin = 0, ymax = 1000,
xtick distance = 1,
ytick distance = 100,
grid = both,
xlabel style={font=\small},        
ylabel style={font=\small, yshift=-3pt},  
tick label style={font=\small},
minor tick num = 1,
width = \columnwidth,
height = 0.4\textwidth,
xlabel = {Range of mean-normalized weights},
ylabel = {Number of Optimal shots},
]
\addplot [color=red, mark=square] table [x = diff_weight, y = optimal_shots, col sep=comma,]{diff_weight_sss_qa_r5_q0.csv};
\addlegendentry{query-1}
\addplot [color=blue, mark=halfsquare*] table [x = diff_weight, y = optimal_shots, col sep=comma,]{diff_weight_sss_qa_r5_q4.csv};
\addlegendentry{query-2}
\addplot [color=black, mark=triangle] table [x = diff_weight, y = optimal_shots, col sep=comma,]{diff_weight_sss_qa_r5_q6.csv};
\addlegendentry{query-3}
\addplot [color=olive, mark=triangle*] table [x = diff_weight, y = optimal_shots, col sep=comma,]{diff_weight_sss_qa_r5_q7.csv};
\addlegendentry{query-4}
\addplot [color=violet, mark=halfsquare right*] table [x = diff_weight, y = optimal_shots, col sep=comma,]{diff_weight_sss_qa_r5_q8.csv};
\addlegendentry{query-5}
\addplot [color=red, mark=diamond] table [x = diff_weight, y = optimal_shots, col sep=comma,]{diff_weight_sss_qa_r5_q53.csv};
\addlegendentry{query-6}
\addplot [color=blue, mark=diamond*] table [x = diff_weight, y = optimal_shots, col sep=comma,]{diff_weight_sss_qa_r5_q159.csv};
\addlegendentry{query-7}
\addplot [color=black, mark=halfdiamond*] table [x = diff_weight, y = optimal_shots, col sep=comma,]{diff_weight_sss_qa_r5_q202.csv};
\addlegendentry{query-8}
\addplot [color=olive, mark=*] table [x = diff_weight, y = optimal_shots, col sep=comma,]{diff_weight_sss_qa_r5_q271.csv};
\addlegendentry{query-9}
\addplot [color=violet, mark=star] table [x = diff_weight, y = optimal_shots, col sep=comma,]{diff_weight_sss_qa_r5_q5.csv};
\addlegendentry{query-10}
\end{axis}
\end{tikzpicture}
\caption{The above figure shows the variation of the number of optimal shots with the range of mean-normalized weights for the experiments performed on D-Wave's quantum process unit. All these experiments have been performed above for the split that consists of eight binary variables in the QUBO search space for the query of five relations. As shown in Appendix~\ref{appendix:sqlqueries}, the SQL queries were used to extract relevant data for the experiments performed using simulated annealing in Figure~\ref {fig:optimal_shots_sa_mean-normalised_weight_range} and on the quantum annealer. For a detailed list of queries, refer to Listing~\ref{lst:sqlqueries}.}
\label{fig:optimal_shots_qa_mean-normalised_weight_range}
\end{figure}

To study the effect of selectivity on solving join ordering using quantum and classical algorithms, we performed experiments with queries consisting of five relations. 
This choice allowed us to analyze the impact of selectivity on QPU performance effectively. 
For queries consisting of fewer than five relations, optimal join order was achieved with near $100\%$ accuracy by employing the SQSS approach.
In contrast, for queries involving joining more relations, such as six, the effectiveness of selectivity on QPU performance diminished. 
Additionally, the increased problem complexity required a greater number of experiment runs and significantly more time to solve the query, whereas five-relation queries were solved in comparatively less time.
When experiments were conducted on meeting the condition of selectivity on the database, followed by solving the QUBO problems using SQSS, the performance of both algorithms, e.g., SA and QA, increased.
In further investigation, with improved algorithm performance, the range of mean-normalized weights assigned to the binary variable in the QUBO formulation decreased, as we can see in figures~\ref{fig:optimal_shots_sa_mean-normalised_weight_range} and \ref{fig:optimal_shots_qa_mean-normalised_weight_range}.
Although comparing the results depicted in both the figures reveals that selectivity does not consistently have a more positive impact on the QA method when compared with the SA method.
This observation suggests that the performance of the QPU may vary even on repeating the same experiments due to other factors that affect the performance of the QPU, leading to differences in the results retrieved in each run.

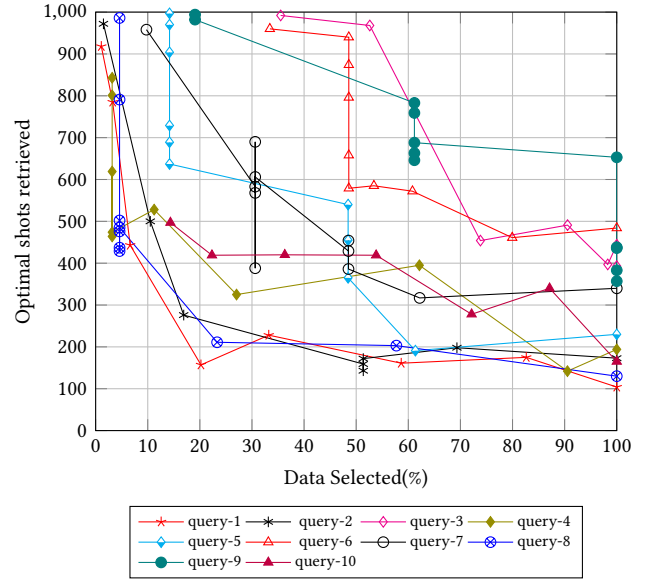
\begin{figure}[htb!]
\centering
\begin{tikzpicture}
\begin{axis}[legend style={nodes={scale=0.7, transform shape},
             at={(0.5,-0.18)}, anchor=north, legend columns= 4},
xmin = 0, xmax = 100,
ymin = 0, ymax = 1000,
xtick distance = 10,
ytick distance = 100,
grid = both,
xlabel style={font=\small},        
ylabel style={font=\small, yshift=-3pt},  
tick label style={font=\small},
minor tick num = 0,
width = \columnwidth,
height = 0.4\textwidth,
xlabel = {Data Selected(\%)},
ylabel = {Optimal shots retrieved},
]
\addplot [color=red, mark=star] table [x = data_perc, y = optimal_shots, col sep=comma,]{data_perc_QA_r5_q0.csv};
\addlegendentry{query-1}
\addplot [color=black, mark=asterisk] table [x = data_perc, y = optimal_shots, col sep=comma,]{data_perc_QA_r5_q4.csv};
\addlegendentry{query-2}
\addplot [color=magenta, mark=diamond] table [x = data_perc, y = optimal_shots, col sep=comma,]{data_perc_QA_r5_q6.csv};
\addlegendentry{query-3}
\addplot [color=olive, mark=diamond*] table [x = data_perc, y = optimal_shots, col sep=comma,]{data_perc_QA_r5_q7.csv};
\addlegendentry{query-4}
\addplot [color=cyan, mark=halfdiamond*] table [x = data_perc, y = optimal_shots, col sep=comma,]{data_perc_QA_r5_q8.csv};
\addlegendentry{query-5}
\addplot [color=red, mark=triangle] table [x = data_perc, y = optimal_shots, col sep=comma,]{data_perc_QA_r5_q53.csv};
\addlegendentry{query-6}
\addplot [color=black, mark=o] table [x = data_perc, y = optimal_shots, col sep=comma,]{data_perc_QA_r5_q159.csv};
\addlegendentry{query-7}
\addplot [color=blue, mark=otimes] table [x = data_perc, y = optimal_shots, col sep=comma,]{data_perc_QA_r5_q202.csv};
\addlegendentry{query-8}
\addplot [color=teal, mark=oplus*] table [x = data_perc, y = optimal_shots, col sep=comma,]{data_perc_QA_r5_q271.csv};
\addlegendentry{query-9}
\addplot [color=purple, mark=triangle*] table [x = data_perc, y = optimal_shots, col sep=comma,]{data_perc_QA_r5_q5.csv};
\addlegendentry{query-10}
\end{axis}
\end{tikzpicture}
\caption{This figure shows the variation in the data selected with enhanced performance of the quantum annealer.} 
\label{fig:reduced_dataset_optsts_QA_sss}
\end{figure}

Referring to figure~\ref{fig:reduced_dataset_optsts_QA_sss}, each successive selectivity condition resulted in changes to the cardinality estimation (or weight distribution), which altered the energy landscape of the QUBO. 
These changes positively influenced the performance of D-Wave’s QPU.
Similarly, SA exhibited a similar trend in performance variation. 

\section{Summary And Conclusion}
\label{sec:conclusions}
Solving combinatorial optimization problems is one of the most promising applications of quantum computing.
With the advent of the first quantum computers, the field of quantum computing has generated a great deal of interest.
It is not surprising to see that several research efforts have tried to apply quantum computing to an important combinatorial optimization problem, such as query optimization problems.
To address some limitations of existing work and further improve the JOO performance of quantum computing, we studied the elimination of the Cartesian product from the query join order plan with described use cases in section~\ref{sec:Utilisation_of_ECP_for_JOO}.
The adoption of the techniques will significantly reduce the number of intermediate join results and shrink the search space of quantum join optimization algorithms.

In section~\ref{subsec:result-analysis}, we extensively evaluate the techniques using various quantum algorithms (QA, SA, QAOA, and VQE) on quantum computing devices such as D-Wave’s QPU and universal quantum simulators.
In the experimental evaluation, we benchmarked our study with real-world SQL queries of varying complexities from small to large, as well as different forms of join graphs.
We investigated the relation of the number of qubits required for quantum annealing to the number of cross-join pairs in the SQL query, the relation of the number of experimental runs required for the queries of various relations, and the variation of the fraction of the optimal shots with the varying number of relations in the query.
The results of the evaluation show that our techniques further improve the JOO performance for all the quantum algorithms we tested.
For queries of higher complexity (e.g., queries including six relations), even after splitting, it may contain more binary variables in the search space in comparison to the split QUBO search space for queries with lower complexity.
Experimenting with simulated annealing shows a significant improvement in retrieving optimal shots on solving QUBO compared with the previous method while using QPU. 
Due to limited access and the limitations of the current state-of-the-art, we retrieved fewer optimal shots than expected.

Furthermore, we conducted a study to analyze how query selectivities influence the performance of quantum and classical optimization algorithms, such as SA and QA. 
We experimentally evaluated the relationship between query selectivities and the number of optimal shots, highlighting their impact on the performance of quantum optimization algorithms.
In addition, we examined how selectivity affects the weight distribution in the optimization process.
Although our experiments were limited to queries involving five relations for the benchmarked dataset, the ErgastF1 dataset. 

For future work, exploring the integration of selectivity with the ECP-SQSS presents a promising avenue to evaluate its impact on the performance of quantum algorithms. 
Additionally, selectivity could be tested on queries with higher complexity to better understand its effectiveness in more challenging scenarios.
With advancements in quantum hardware, such as the recently developed quantum annealers based on new topologies like Zephyr, it would be fascinating to investigate their performance on our proposed method for higher-complexity queries. 
Scalability will play a crucial role in addressing the query optimization problem, and the development of more advanced quantum devices has the potential to further reduce the experimental runs required, paving the way for more efficient quantum-based solutions.

\section*{Acknowledgements} This work is funded by the German Federal Ministry of Education and Research within the funding program "Quantum Technologies—From Basic Research to Market," contract number 13N16090.

\section*{Code Availability}

The source code and experimental artifacts used in this study are publicly available at:
\url{https://anonymous.4open.science/r/Improved_Join_Order_Optimization-ECP-SQSS--8900}

Experiments involving quantum hardware require access to external cloud-based quantum computing platforms. All necessary parameters and configurations are included to facilitate reproducibility.

\bibliographystyle{ACM-Reference-Format}
\bibliography{main}

\clearpage
\appendix
\section{Appendix}
\label{appendix:sqlqueries}

This appendix contains the SQL queries used in our experiments for selectivity as shown in Figure \ref{fig:optimal_shots_qa_mean-normalised_weight_range}.

\begin{lstlisting}[language=SQL, caption={SQL Queries for Experimental Analysis}, label={lst:sqlqueries}]
-- Query - 01
SELECT * FROM
circuits,
 constructorresults,
 constructors,
 constructorstandings,
 races
WHERE
constructorresults.constructorId=constructors.constructorId
AND constructorresults.raceId=races.raceId
AND constructorstandings.constructorId=constructors.constructorId
AND constructorstandings.raceId=races.raceId
AND races.circuitId=circuits.circuitId;

-- Query - 02
SELECT * FROM
circuits,
 constructorresults,
 constructors,
 qualifying,
 races
WHERE
constructorresults.constructorId=constructors.constructorId
AND constructorresults.raceId=races.raceId
AND qualifying.constructorId=constructors.constructorId
AND qualifying.raceId=races.raceId
AND races.circuitId=circuits.circuitId;

-- Query - 03
SELECT * FROM
circuits,
 constructorresults,
 constructors,
 races,
 seasons
WHERE
constructorresults.constructorId=constructors.constructorId
AND constructorresults.raceId=races.raceId
AND races.circuitId=circuits.circuitId
AND races.year=seasons.year;

-- Query - 04
SELECT * FROM
circuits,
 constructorresults,
 constructorstandings,
 driverstandings,
 races
WHERE
constructorresults.raceId=races.raceId
AND constructorstandings.raceId=races.raceId
AND driverstandings.raceId=races.raceId
AND races.circuitId=circuits.circuitId;

-- Query - 05
SELECT * FROM
circuits,
 constructorresults,
 constructorstandings,
 laptimes,
 races
WHERE
constructorresults.raceId=races.raceId
AND constructorstandings.raceId=races.raceId
AND laptimes.raceId=races.raceId
AND races.circuitId=circuits.circuitId;

-- Query - 06
SELECT * FROM
circuits,
 constructorstandings,
 drivers,
 laptimes,
 races
WHERE
constructorstandings.raceId=races.raceId
AND laptimes.driverId=drivers.driverId
AND laptimes.raceId=races.raceId
AND races.circuitId=circuits.circuitId;

-- Query - 07
SELECT * FROM
constructorresults,
 constructorstandings,
 driverstandings,
 laptimes,
 races
WHERE
constructorresults.raceId=races.raceId
AND constructorstandings.raceId=races.raceId
AND driverstandings.raceId=races.raceId
AND laptimes.raceId=races.raceId;

-- Query - 08
SELECT * FROM
constructorresults,
 laptimes,
 pitstops,
 qualifying,
 races
WHERE
constructorresults.raceId=races.raceId
AND laptimes.raceId=races.raceId
AND pitstops.raceId=races.raceId
AND qualifying.raceId=races.raceId;

-- Query - 09
SELECT * FROM
constructors,
 driverstandings,
 pitstops,
 qualifying,
 races
WHERE
driverstandings.raceId=races.raceId
AND pitstops.raceId=races.raceId
AND qualifying.constructorId=constructors.constructorId
AND qualifying.raceId=races.raceId;

-- Query - 10
SELECT * FROM
circuits,
 constructorresults,
 constructors,
 races,
 results
WHERE
constructorresults.constructorId=constructors.constructorId
AND constructorresults.raceId=races.raceId
AND races.circuitId=circuits.circuitId
AND results.constructorId=constructors.constructorId
AND results.raceId=races.raceId;
\end{lstlisting}
\end{document}